\documentclass[useAMS,usenatbib]{mn2e}
\usepackage{textcomp,enumitem}
\usepackage{graphicx,natbib}	% Including figure files
\usepackage{amsmath}	% Advanced maths commands
\usepackage{amssymb}	% Extra maths symbols
\usepackage{multicol}
\usepackage{multirow}% Multi-column entries in tables
\usepackage{bm}
\usepackage{pdflscape}	% Landscape pages
\usepackage{float}
\usepackage[colorlinks=true,citecolor=blue]{hyperref}
\usepackage{longtable}
\usepackage{xcolor}
\usepackage{ulem}
\usepackage{supertabular}
\usepackage{array}

\newcommand{\apj}{ApJ}
\newcommand{\apjs}{ApJS}
\newcommand{\apjl}{ApJL}
\newcommand{\aap}{A{\&}A}
\newcommand{\aaps}{A{\&}AS}
\newcommand{\mnras}{MNRAS}
\newcommand{\aj}{AJ}
\newcommand{\araa}{ARA\&A}
\newcommand{\pasp}{PASP}

\newcommand{\nat}{Nature}

\newcommand{\apss}{apss}
\newcommand{\fcp}{fcp}

\title[UV emission in Dual Nuclei Galaxies ]
{Study of Star-formation in Dual Nuclei Galaxies using UVIT observations}
\author[Rubinur et al.]{{K. Rubinur}$^{1,2}$\thanks{E-mail: rubinur.khatun@astro.uio.no}, {M. Das}$^3$, {P. Kharb}$^2$, {J. Yadav}$^{3,4}$ {C. Mondal}$^5$, {P.T. Rahna}$^{6,7}$ \\
$^1$ Institute of Theoretical Astrophysics, University of Oslo, P.O box 1029 Blindern, 0315 OSLO, Norway\\
$^2$ National Centre for Radio Astrophysics - Tata Institute of Fundamental Research (NCRA-TIFR), S. P. Pune University Campus,\\ Ganeshkhind, Pune 411007, India\\
$^3$ Indian Institute of Astrophysics, Koramangala II Block, Bangalore 560034, India\\
$^{4}$Pondicherry University, R.V. Nagar, Kalapet, 605014, Puducherry, India \\
$^5$ Inter-University Centre for Astronomy and Astrophysics, Ganeshkhind, Post Bag 4, Pune 411007, India\\
$^6$ CAS Key Laboratory for Research in Galaxies and Cosmology, Shanghai Astronomical Observatory, \\Shanghai, 200030, China\\
$^7$ Centro de Estudios de F\'isica del Cosmos de Arag\'on (CEFCA), Plaza San Juan 1, E-44001 Teruel, Spain}

\begin{document}
\label{firstpage}
\maketitle

\begin{abstract}
We have used the Ultraviolet Imaging Telescope (UVIT) aboard AstroSat to study star formation in a sample of nine dual nuclei galaxies with separations $\loa$ 11 kpc, which is an expected outcome of galaxy mergers. To minimize the contribution of active galactic nuclei (AGN) emission, we have used mid-IR color cut-offs and masked the AGN-dominated nuclei. The UV continuum slope ($\beta$) is used to calculate the internal extinction (A$_V$) which ranges from 0.53 to 4.04 in the FUV band and 0.44 to 3.10 in the NUV band for the sample. We have detected $1-20$ star-forming clumps (SFCs) in our sample galaxies. The extinction-corrected total FUV star-formation rate (SFR) ranges from $\sim$0.35 to $\sim$32 M$_\odot$ yr$^{-1}$. Our analyses of A$_V$, specific SFR (sSFR) show that dual nuclei sources are associated with dusty, star-forming galaxies. The FUV$-$NUV color maps show redder color in the nuclear and disk regions while bluer color is observed in the outskirts of most galaxies which can be due to embedded dust or different stellar populations. We have found some signatures of possible stellar/AGN feedback like a ring of star formation, a redder ring around blue nuclei, etc. However, further observations are required to confirm this.

\end{abstract}

\section{Introduction} \label{intro}
Galaxies show a bimodality in the star-formation rate (SFR) - stellar mass (M$_\star$) plane. It has been found that while early-type galaxies, which are mostly elliptical and S0 galaxies with old stellar populations, form the red cloud, the late-type spiral galaxies, which are star-forming, occupy the blue cloud. The green valley is defined as the region of transition between early-type and late-type galaxies \citep{daddi2007,salim2007,elbaz2018}. Statistically large samples of early-type galaxies show that most of the stellar mass has been accumulated in the past 8 billion years \citep{brwon2007}. There are several processes that can help to build the stellar mass: one of the most important being galaxy mergers \citep{hopkins2010}.  

Hierarchical galaxy formation theories predict that galaxies have formed through several major (mass ratio $\geq$ 1:3) and minor (mass ratio $\leq$ 1:3) mergers \citep{volonteri2003, springel2005}. Two spiral galaxies can form an elliptical galaxy through a major merger and it can turn into a star-burst galaxy depending on the availability of gas. Hence, galaxy mergers are the key drivers of galaxy evolution \citep[e.g.,][]{barnes1992}, and understanding these systems is important. The tidal forces due to the interaction produce non-axisymmetric gravitational forces across the disks causing enormous changes in the potential of the galaxies \citep{Bournaud2010a}. These changes cause an increase in cloud collisions and shocks resulting in star-formation (SF) \citep[e.g.,][]{kennicutt1987,saitoh2009,ellison2013}. Therefore, mergers can disturb the galaxies leading to gas inflow towards the galaxy nuclei as well as within the parent galaxies, often leading to starburst activity in the galaxies \citep{hopkins2009}. Observations of such merger-induced SF suggest that most galaxy mergers should go through the star-burst phase \citep{schweizer2005}. However, recent observations show that such star-burst activity is found only in a minority of galaxy mergers. On the other hand, simulations of merging galaxies have shown that SFR is increased when galaxies are close to each other during the first, second pericenter passes, and finally during coalescence \citep[e.g.,][]{hopkins2006, rupke2010}. However, in between these periods of close separation, which forms most of the interaction time, the SFR increases at most by a factor of two which is much lower than the SFRs expected from star-burst galaxies \citep{moreno2019}.

Several studies have tried to understand the effect of mergers at different galaxy separations, different redshifts, and with different types of galaxies \citep{ellison2013, knapen2015}. Major mergers can grow 20\% of the mass for massive galaxies in z$<$1 which is significant but not sufficient \citep{lopez2010}. This leads to the requirement of minor mergers for galaxy growth. It has been found that minor mergers contribute to almost $\sim$35 \% of the star-formation over the cosmic time \citep{kaviraj2014}. Hence, both major and minor mergers are important to understand the galaxy evolution in terms of star formation and nuclear activity. 

Star formation can be studied using UV and H$\alpha$ observations \citep[e.g.,][]{kennicutt1998, calzetti2013}. The H$\alpha$ arises from massive O and B-type stars. The lifetime of these stars is only $\sim10^{6} - 10^{7}$ years; therefore, H$\alpha$ effectively traces SF only for a short period. On the other hand, UV emission arises from the ionizing radiation of O, B, and A-type stars, as well as some evolved stars, so it traces SF for $\sim10^{6} -10^{8}$ years, i.e. 10 times longer than H$\alpha$. A part of UV emission in the galaxies is absorbed by the dust within the galaxy and is re-emitted in the infrared (IR). One needs to correct the UV emission for dust extinction to calculate the total SFR. 

The gas inflow \citep{hopkins2009} towards the centers during the merger can ignite the accretion activity to the central supermassive black hole and turn them into active galactic nuclei (AGN) \citep[mass $\sim10^{6-8}~M_{\sun}$;][]{mihos1996,mayer2007}. Studies have found that the AGN fraction increases with the number of mergers \citep{ellison2011}. Once AGN activity is triggered and the SMBHs reach a certain critical mass \citep{Ishibashi2012}, they give out energy to the surrounding medium via winds, jets, and radiation. The winds can trigger star formation beyond the AGN by shocking gas; the outflowing winds can also suppress gas infall due to the pressure of the gas and the radiation. This is collectively called AGN feedback \citep[see][for review]{fabian2012, morganti2017, Harrison2017}. In the low mass starburst galaxies, stellar-driven galactic winds are also significant \citep[see][for review]{zhang2018}. Stellar-driven feedback can work together with AGN feedback in some galaxies \citep{rupke2011}. Theories show that during mergers, after an intense burst of star formation and black hole accretion, the feedback processes can remove the gas as well as heat the interstellar medium (ISM) which prevents gas cooling. This process can suppress the star formation and transform blue galaxies into red galaxies in SFR-M$_\star$ plane \citep[][and reference therein]{hopkins2008, ellison2022}.

The whole process of star-formations, and AGN activity during mergers is not yet well understood, especially in dual nuclei galaxies which are one of the outcomes of galaxy mergers. Many dual nuclei galaxies are found in the surveys of interacting galaxies such as ultra-luminous infrared galaxies \citep[ULIRGs;][]{mazzarella1991}. \citet{mazzarella1988, mazzarella2012} have studied individual systems using multi-wavelength observations. Some studies approach dual nuclei galaxies from the merger hypothesis angle \citep{gemino2004,mezcua2014} while others focus on the nuclear emission to detect AGN pairs (or dual AGN) \citep{koss2012, rubinur2021}. Thus, dual nuclei galaxies can help us to investigate the final stages of galaxy mergers.

So far, most of the studies in the literature have explored star formation in galaxies as well as galaxy pairs or mergers using GALEX\footnote{http://www.galex.caltech.edu/} UV observations which has an angular resolution of $\sim$ 5$\arcsec$ \citep[e.g.,][]{smith2010, yuan2012}. However, the Ultra-violet imaging telescope (UVIT) \citep{kumar2012} onboard  ASTROSAT, has a better angular resolution of $\sim 1.2\arcsec$. Several nearby galaxies have been explored which have produced better-resolution images. Some examples of such studies are star-forming clumps in extended ultraviolet (XUV) disk galaxies \citep[e.g.,][]{yadav2021, das.etal.2021}, dwarf galaxies \citep[e.g.,][]{mondal2018}, spiral galaxies \citep[e.g.,][]{rahna2018}, post-merger galaxies \citep[e.g.,][]{koshy2018, george2018a,yadav.etal.2023}. %Some of these studies will be discussed in the later part of this paper.

In this paper, we have studied the star formation in a sample of UV-bright dual nuclei galaxies using UVIT observations. The paper is structured as follows: the sample selection is discussed in section 2. Section 3 describes the UVIT observations, details of archival IR data, and data analysis. Further image analysis and estimation of required parameters like extinction, SFR, M$_\star$ are discussed in section 4. The results are presented in section 5 along with the discussions. The summary and conclusion are given in section 6. We have used the cosmology with $\Omega_{m}=~0.27$, and $H_{0}=~73.0$~km~s$^{-1}$~Mpc$^{-1}$. The spectral index, $\alpha$, is defined such that the flux density at frequency $\nu$ is S$_\alpha\propto~\nu^\alpha$

\section{Sample selection} \label{sample_selection}
We started this as a pilot study to explore star formation in dual nuclei galaxies with UVIT. Here, we define dual nuclei as those with a projected separation of $\sim$ 10 kpc and we include nuclei of all types, i.e. both AGN and SF types. Our main criterion is that the nuclei should be embedded in one common envelope or closely interacting. For the UVIT observations, the sample had to go through several instrument/UV criteria such as (a) galaxies with strong UV detection in previous UV surveys such as in GALEX\footnote{https://galex.stsci.edu/GR6/?page=mastform}, (b) the sources must be visible in the sky (tool {\it Astroviewer}\footnote{http://issdc.gov.in/astroviewer/index.html}), (c) there should not be any bright source in the field which could harm the telescopes (tool {\it BSWT}\footnote{https://uvit.iiap.res.in/Software/bswt}). We started from an initial sample of merger systems from \citet{Mezcua2011}; they carried out a photometric study of a sample of 52 dual nuclei systems. We selected six systems from their study. To increase the sample, we included one sample galaxy from \citet{ge2012} and two sample galaxies from \citet{liu2011a}, both of which are studies of narrow emission line galaxies with double-peaked AGN (DPAGN) that are dual AGN candidates \citep{rubinur2019}. Along with these nine sources, we have included another source  ESO509-IG006 which is a closely interacting galaxy pair with a separation of $\sim$11 kpc \citep{Guainazzi2005}. Our final sample had 10 dual nuclei galaxies. One of these galaxies (MRK 212) is presented in \citet{rubinur2021}, where two SF knots near one of the nuclei are detected in the 15 ksec UVIT image. Hence, the rest of the nine galaxies (Table \ref{sample}) are presented in this work.

%\citet{ge2012} compiled a sample of  3030 narrow emission line galaxies that show double-peaked AGN (DPAGN) which are dual AGN (DAGN) candidates. They separated the galaxies into different categories like Type I, Type II, SF, and double nuclei depending on line ratios and morphologies. We have studied a sub-sample of 20 DPAGN from \citet{ge2012} and explored with high-resolution radio observations to detect DAGN \citep{rubinur2017, rubinur2019}.  \citet{liu2011a} studied 1286 AGN pairs at z $\sim$ 0.1 with line of velocity offsets $<$ 600 km s$^{-1}$ and projected separation $<$ 100 h$_{70}^{-1}$ kpc. Their work increases the number of AGN pairs in those scales by an order of magnitude. Along with these nine sources, we have included another source  ESO509-IG006 which is a closely interacting galaxy pair with a separation of $\sim$11 kpc. 

\begin{figure}
\includegraphics[width=1\columnwidth, trim= 3cm 0 0 0]{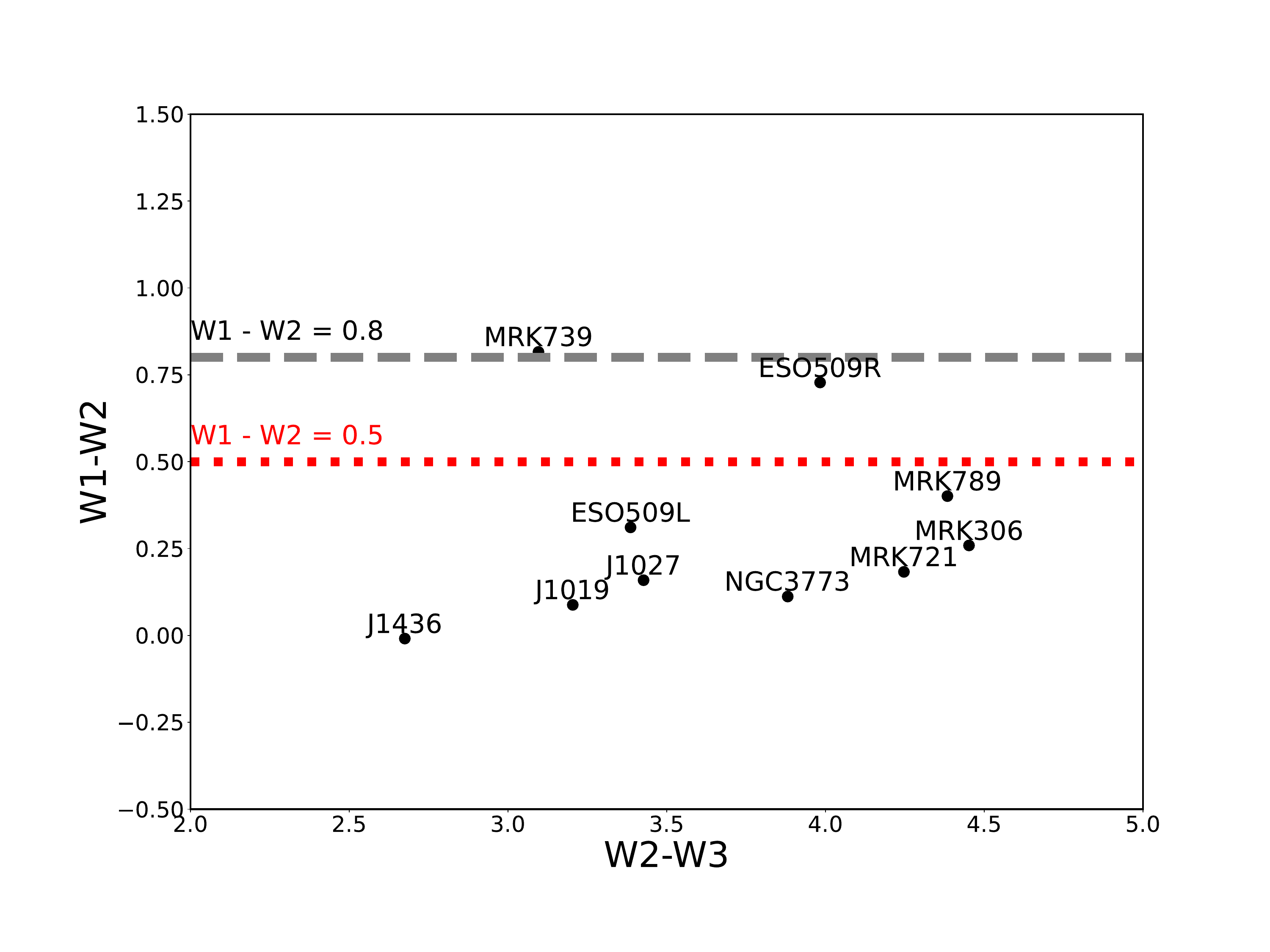}
\caption{\small Wise color - color plot using W1-W2 (mag) vs W2-W3 (mag). The sample galaxies are plotted as black dots. Several color cut-offs are present to separate the AGN emission from star formation. We have used two such cut-offs: W1-W2 = 0.8 for single AGN \citep{stern2012} while W1 - W2 = 0.5 for multiple AGN  \citep{blecha2018}. The dual AGN MRK 739 and the western nuclei of ESO509-IG066 fall above the limit.}
\label{wise_color}
\end{figure}

\begin{figure*}
\centering
\includegraphics[width=18cm, trim = 0 0 0 0]{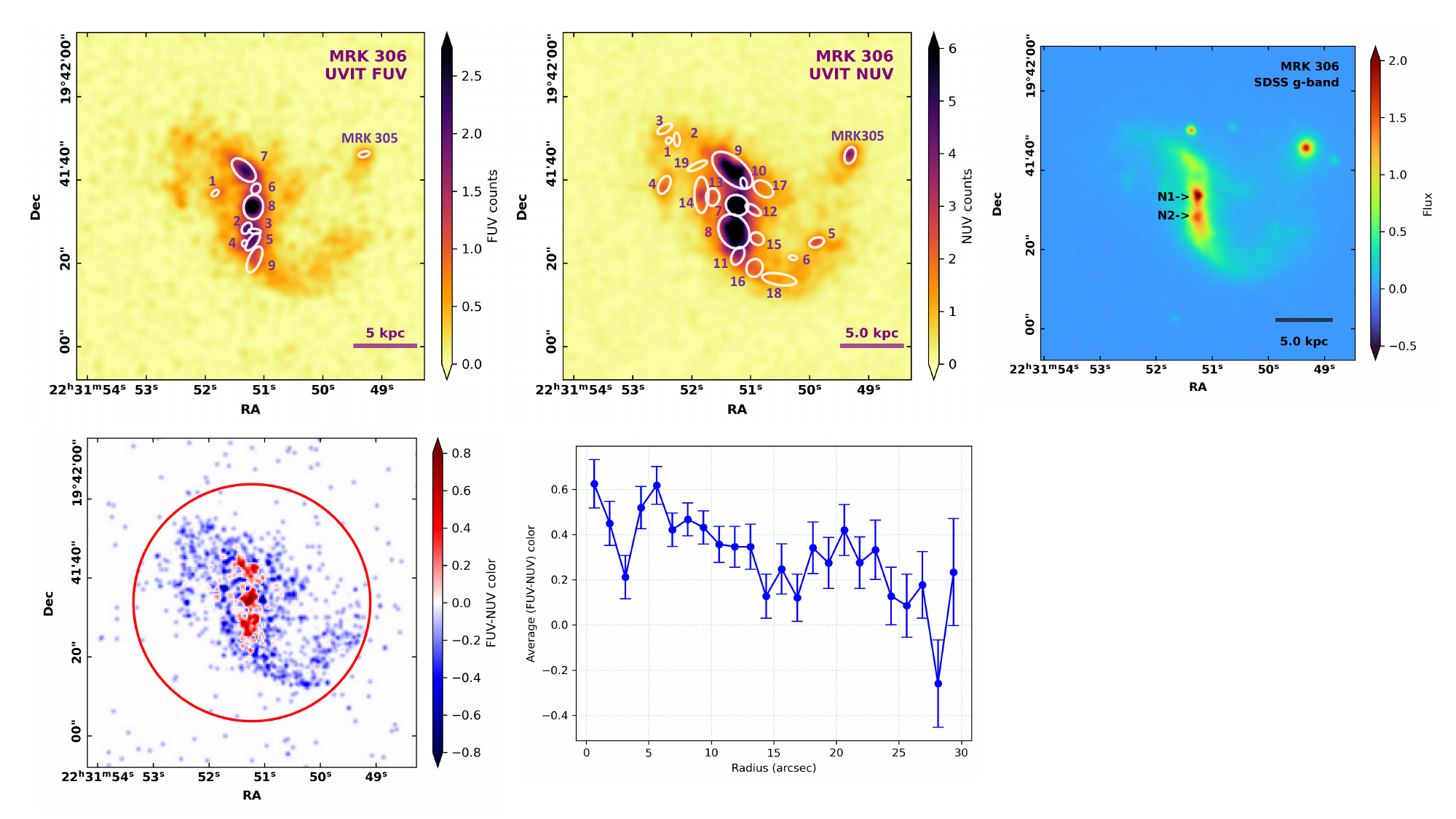}
\caption{\small Multi-band images of MRK 306 along with the detected star-forming clumps. {\it Upper left}: The UVIT FUV image; {\it upper middle:} The UVIT NUV image; {\it upper right:} the SDSS g-band image indicating the dual nuclei; {\it lower left:} The FUV-NUV color map; {\it lower middle:} The radial profile of FUV-NUV color. This is obtained with the annulus (width = 3 pixels) starting from the center till the red circle as shown on the color map.
The neighbor galaxy MRK 305 is masked while making the color map and color profile.}
\label{fig306}
\end{figure*}

\begin{figure*}
\centering

\includegraphics[width=18cm, trim = 0 0 0 0]{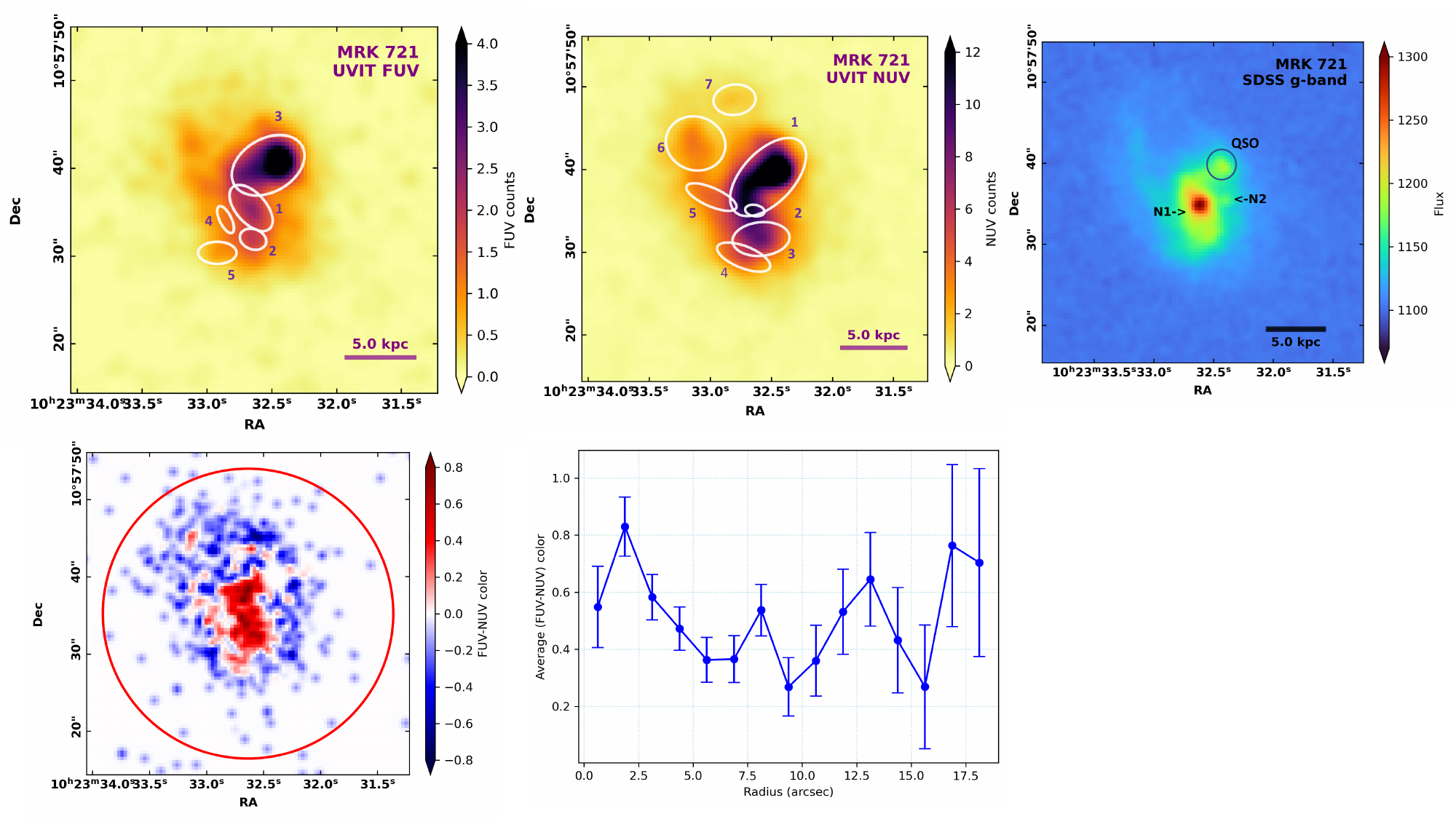}
\caption{\small Multi-band images of MRK 721. Same as figure \ref{fig306}. SFC no 3 in the FUV and SFC no 1 in the NUV images involve a bright source on the northwest which is assigned as a QSO source as shown in the SDSS image ({\it upper right}) with a black circle. While performing the photometry as well as the color map, and color profile, we masked the QSO with an aperture radius of 5 pixels.}
\label{fig721}
\end{figure*}

\begin{figure*}
\centering
\includegraphics[width=20cm, trim = 0 0 0 0]{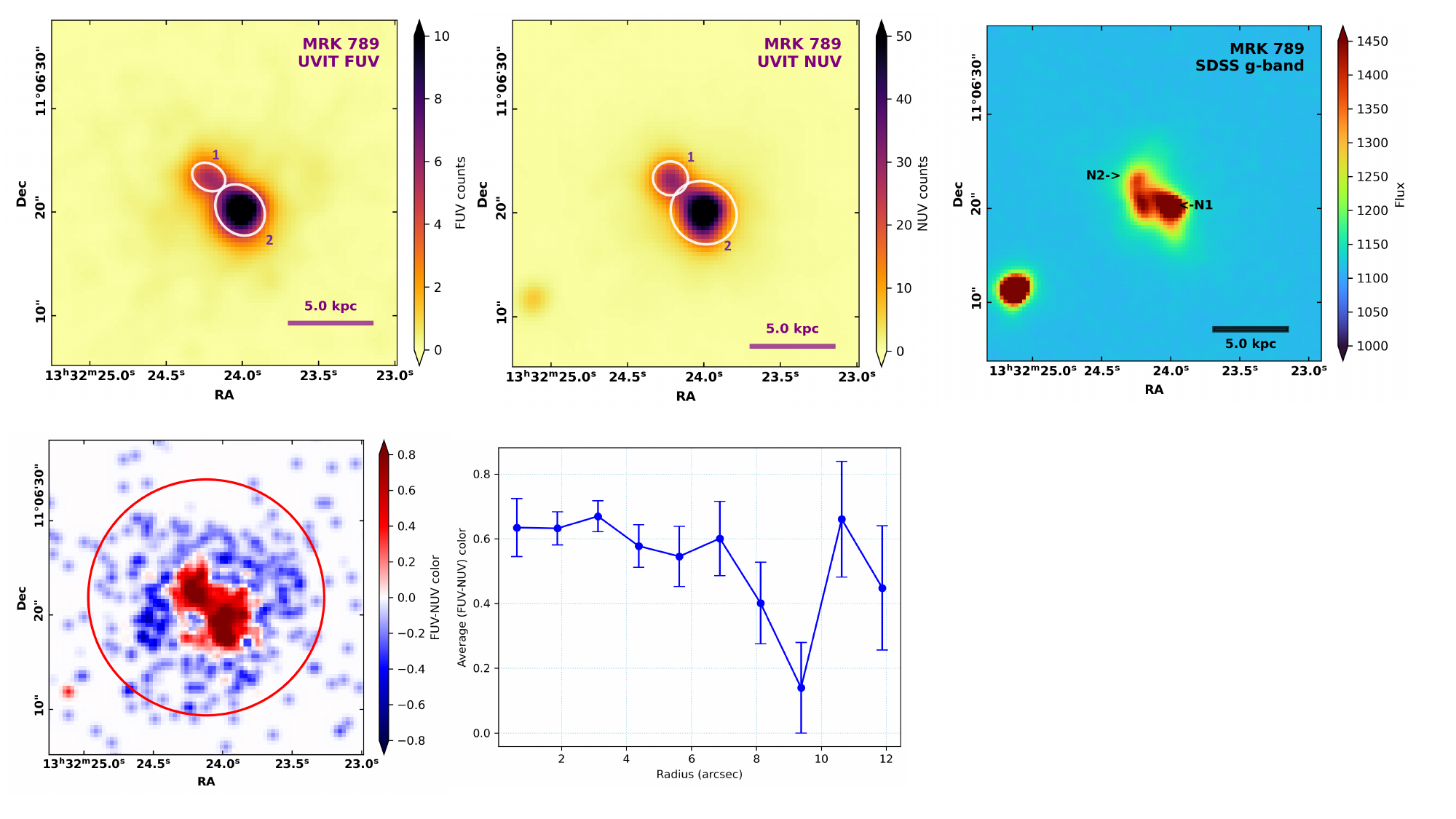}
\caption{\small Multi-band images of MRK 789. Same as figure \ref{fig306}.}
\label{fig789}
\end{figure*}

\begin{figure*}
\centering
\includegraphics[width=18cm, trim = 0 0 0 0]{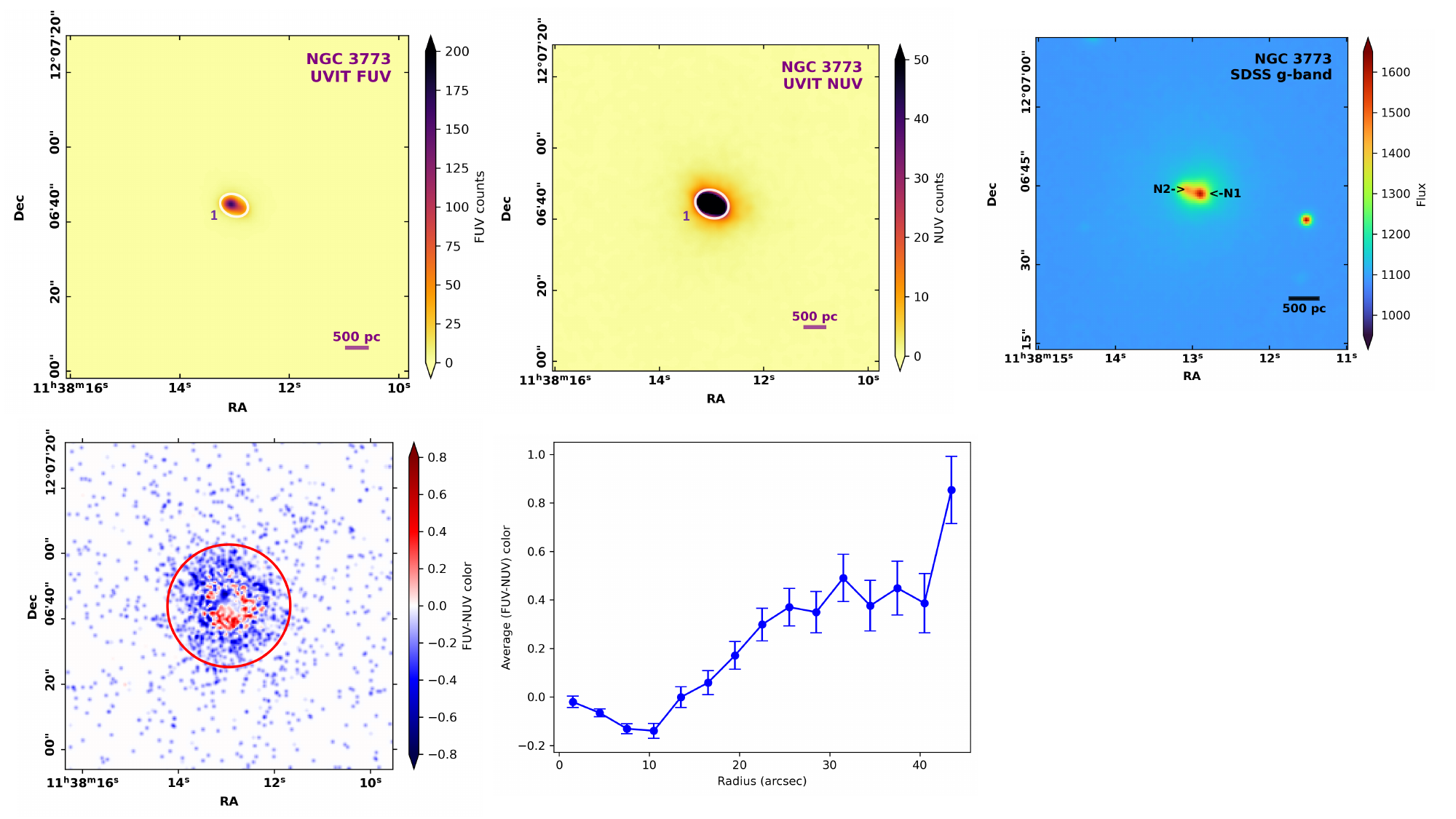}
\caption{\small Multi-band images of NGC 3773.  Same as figure \ref{fig306}. The radial profile of FUV-NUV color shows a blue color in the nuclear region surrounded by redder emission which is opposite in the rest of the galaxies.}

% Images of the dual AGN galaxy Mrk739. (i) The FUV image has detected both nuclei as well as the star-forming clumps in the galaxy disk. (ii) The optical g-band image overlaid with the radio 1.4 GHz FIRST contours. (iii) The 2MASS K-band images has detected two stellar bulges.}
\label{fig3773}
\end{figure*}

\begin{figure*}
\centering
\includegraphics[width=18cm, trim = 0 0 0 0]{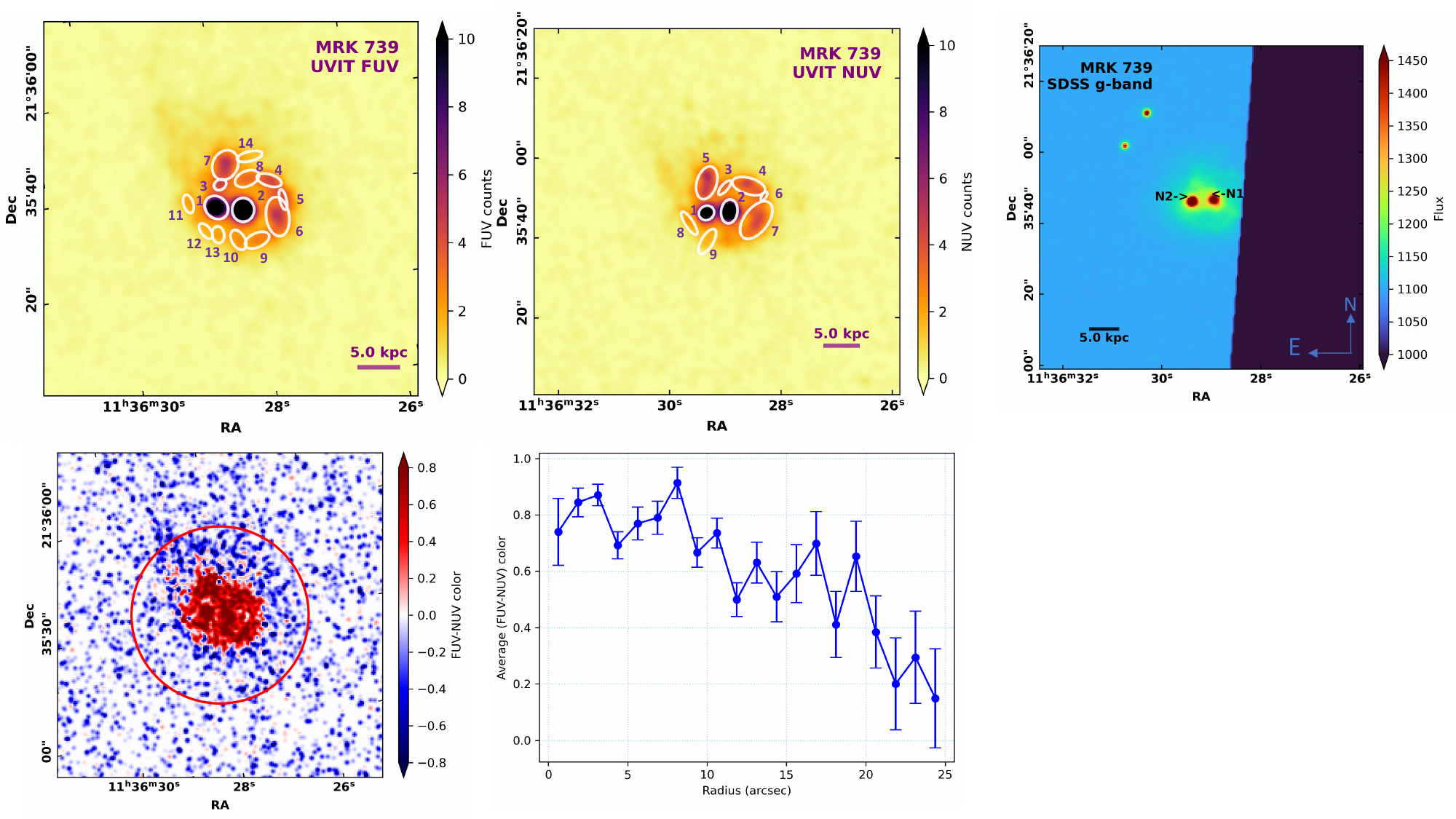}
\caption{\small Multi-band images of MRK 739.  Same as figure \ref{fig306}. This hosts a dual pair of AGNs (SFC 1, 2). We have masked both the nuclear region with a 3-pixel aperture radius and then the photometry is done. The directions are shown with the arrows in the lower right corner of the optical SDSS image. SFC$_{FUV}$ 3, 7 and SFC$_{NUV}$ 5 are discussed further to explore AGN feedback effect in section \ref{feedback}.}
\label{fig739}
\end{figure*}

\begin{figure*}
\centering
\includegraphics[width=18cm, trim = 0 0 0 0]{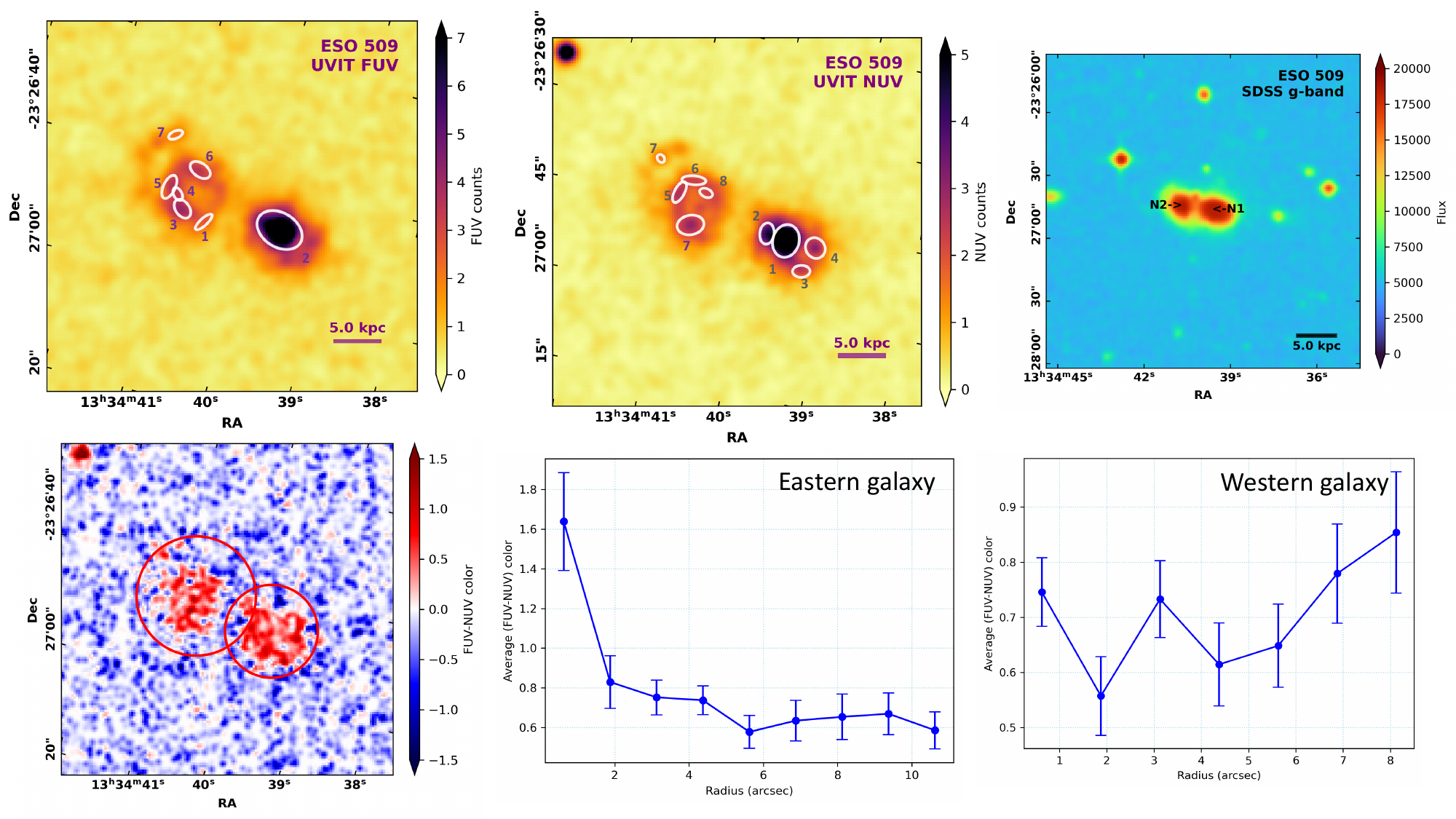}
\caption{\small Multi-band images of ESO509-IG066. Same as figure \ref{fig306}; as the parents' galaxies are yet separated, those are treated as individuals in radial profile analyses and shown in the lower middle and lower right plots. The western nucleus was masked with an aperture of a 3-pixel radius while performing photometry. The eastern nucleus (left) is not detected in FUV image and is surrounded by SFC no 1, 3, 4, 5, and 6 in a ring shape (see section \ref{feedback}). The direction of the images are same as figure \ref{fig739}.}
\label{figESO}
\end{figure*}

\begin{figure*}
\centering
\includegraphics[width=18cm, trim = 0 0 0 0]{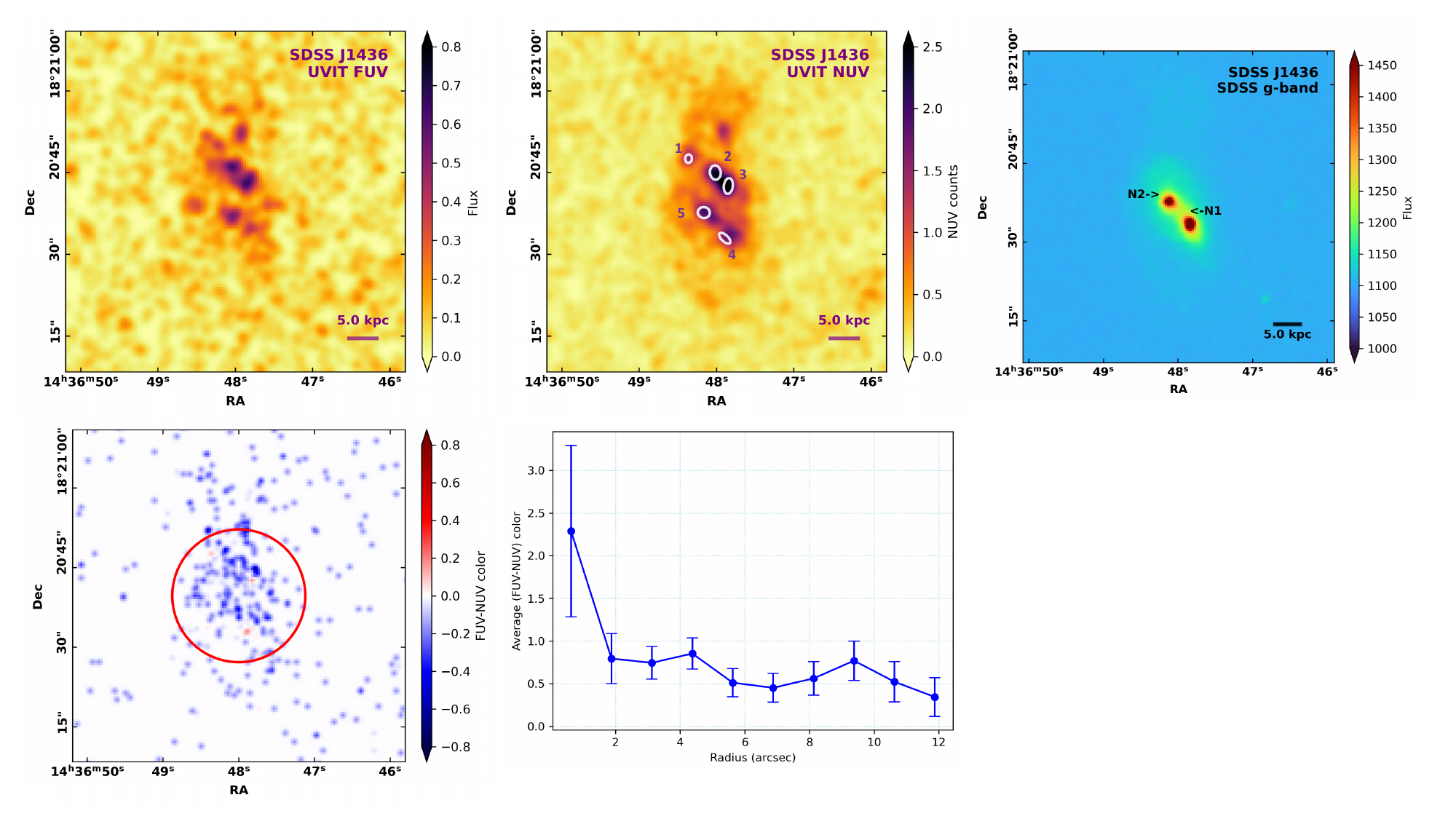}
\caption{\small Multi-band images of SDSS J1436. Same as figure \ref{fig306}. {\it SExtractor} could not detect any SFC in the FUV image. 
}
\label{fig1436}
\end{figure*}

\section{Observations and data reduction}
\subsection{UV}
Our primary motivation in this study is to understand global star formation in galaxies as well as local star formation traced by the star-forming clumps using high-resolution UV images. The main instrument used in this study is the UVIT, which is one of the five payloads onboard India's first Astronomical satellite AstroSat \citep{kumar2012}. The UVIT consists of two co-aligned Ritchey Chretien UV telescopes with a field of view of $28\arcmin$. One telescope is assigned for FUV observations (1300 - 1800 \AA) and the other one for the NUV (2000 - 3000 \AA) and optical bands. The expected spatial resolution of the telescopes is $\sim1.2\arcsec$ to $\sim1.5\arcsec$, which is more than three times better than GALEX ($\sim5\arcsec$). The individual bands have multiple filters with different bandwidths. 

We obtained UVIT data for nine of these galaxies in the initial cycles (A02$-$A04). After inspection of the initial short exposure ($1-5$ ksec) observations, we obtained deep observations ($\geq6$ ksec) for two of these galaxies in cycle A07 (Table \ref{uvit_obs}). The initial observations were carried out with both the NUV and FUV filters (A02$-$165 and A03$-$091). However, the NUV channel stopped working since A04. So, here we present UVIT observations of nine dual nuclei galaxies where eight have both FUV and NUV band data while one object has only FUV data. We have used GALEX images and magnitudes wherever needed. The details of the observations are given in Table \ref{uvit_obs}.

The level 1 UVIT data of the sample galaxies were downloaded from the Indian Space Science Data Centre (ISSDC). A graphical user interface CCDLAB \citep{postma2017} was used to reduce the level 1 data. The CCDLAB does the field distortion and drift corrections. Astrometry for all the sources was done using the GAIA dataset. A tool in CCDLAB matches sources from GAIA catalog \citep{gaia2023} with UVIT sources and applies astrometric corrections.

\subsection{ Infrared}

The mid-IR (MIR) color-color plot (Figure \ref{wise_color}) is a good tool to understand the AGN dominance in galaxy emission. Also, the MIR color can be used to calculate the stellar mass of the galaxies \citep{cluver2014}. In our study,  we have used data from the Wide-field Infrared Survey Explorer \citep[WISE;][]{wright2010}. WISE mapped the entire sky at four mid-infrared filters centered at  
W1: 3.4 $\mu$m; W2: 4.6$\mu$m; W3: 12$\mu$m and W4: 22$\mu$m. The photometry magnitudes are taken from IRSA page\footnote{https://irsa.ipac.caltech.edu/frontpage/} and used for further analysis.  

\section{Analysis}
In this section, we discuss the steps taken for further UV image analysis as well as calculating the extinction, star-formation rate, and stellar masses. 

\subsection{AGN dominance and masking}
Five out of the nine galaxies have confirmed AGN (or multiple AGN) (Table \ref{sample}). It is therefore important to understand the contribution of AGN activity to the galaxy UV emission. The AGN spectra are expected to be redder than the non-active galaxy spectra in the 1-10 $\mu$m range. Hence, the AGN host galaxies have a different location in the mid-IR color-color diagram. Several cut-offs are used to segregate the AGN emission \citep{stern2005, jarrett2011}. We have used a standard diagnostic WISE color-color (W2-W3 vs W1-W2) diagram (Figure \ref{wise_color}) to understand if our sample galaxies have any dominant AGN emission or not, using cut-offs from \citet{stern2012} and \citet{blecha2018}. We found that only MRK 739 and the western galaxy of the ESO 509-IG066 system are above the W1 - W2 cutoff of 0.5. As the AGN is not resolved in our images, we have created an aperture with $\sim$ PSF size with a radius of $\sim$ 3 pixels and then masked both the nuclei of MRK 739 and the western nucleus of ESO 509-IG066 using {\it photutlies} library in Python. Later on, we also masked the nuclei of SDSS J101920.83+490701.2 (see section \ref{global_SFR}). It is possible that the nuclei of other two galaxies with AGN (e.g., MRK 789 or SDSS J102700.40+174901.0) have contributed to the UV emission. However, those are not AGN dominant and the nuclei are not resolved enough to perform the mask analysis.

Apart from AGN nuclei masking, we have used this technique on the following galaxies: (i) MRK 306: the companion galaxy MRK 305 is masked while making the color map and radial profile (Figure \ref{fig306}: lower panel), (ii) MRK 721: it is found that a QSO at redshift 0.745 (Figure \ref{fig721}, upper right), coincides with one of the spiral arms of our sample galaxy MRK 721 (redshift 0.032, Table \ref{sample}). We masked this with an aperture of radius 5 pixels as estimated from the SDSS image while performing photometry and making the color map/profile (Figure \ref{fig721}).

\subsection{UV image analysis}
\subsubsection{Identification of clumps} \label{id_SFCs}
To identify the star-forming clumps (SFCs), we have used the Source Extraction and Photometry \citep[SExtractor;][]{bertin1996} library from Python. {\it SExtractor} can perform tasks like background subtraction, source detection and deblending on the fits format data files. The foreground and background sources were masked during this process. Initially, we set parameters like the threshold, which is the minimum count over which the source is defined, minimum area which is the minimum source area that has to be more than the PSF of the telescope to define a source and deblending count which is used so that sources will not blend with each other. We set the threshold as 5$\sigma$, the minimum area as 10, and the deblending count as 0.0005. 
Therefore, the identification of SFCs involves a three-step process. Initially, all pixels exhibiting a signal-to-noise ratio (SNR) of 5$\sigma$ or higher are chosen. Subsequently, a second criterion, which mandates a minimum contiguous area of 10 pixels, is applied. This 10-pixel threshold is implemented to prevent the detection of SFCs smaller than the UVIT PSF, ensuring confident detections. Finally, a deblending ratio based on contrast separates distinct clumps. These parameter choices result in reliable detections and are consistent with those used in related studies involving the identification of SFCs in UVIT images \citep{yadav.etal.2021, nandi2023}. We have detected SFCs separately in FUV and NUV images and hence the clumps (as well as clump IDs) are different in FUV and NUV images (as shown in Figures \ref{fig306} to \ref{fig1027}). In a few cases, multiple clumps in one band get deblended as a single clump in another band. For example, in ESO509-IC066 (Figure \ref{figESO}), two NUV SFCs (id 1, 2) get deblended as one SFC (id 2) in the FUV image. The central region of ESO509-IC066 in the FUV image is exceptionally bright. This situation presents two possible scenarios, either the contrast within this region does not decrease below the given deblending count, or if the contrast does decrease, no more than 10 connected pixels surpass the specified threshold, thus making it challenging for SExtractor to identify separate clumps. 

The images have been smoothed using a 2-pixel Gaussian Kernel, and contrast settings have been adjusted for improved visualization. Consequently, some regions may appear as clumps (e.g., SDSS J1436 FUV), while others may seem relatively sparse (e.g., MRK 721). It is important to emphasize that the unsmoothed images were used to identify SFCs with SExtractor. In the appendix (Figure \ref{unsmoothed_fig}), we provide the unsmoothed FUV image of SDSS J1436 to illustrate why SFCs are not detected there and an image of MRK 721, where the clumps appear relatively distinct with the given criteria, in our sample galaxies. {\it SExtractor} could detect 5 to 20 SFCs in some galaxies, but some of them have only 1 to 3 SFCs (Figure \ref{fig306} - \ref{fig1027}).

\subsubsection{Aperture photometry}
Aperture photometry is carried out to calculate the total counts of the galaxies as well as the counts in the detected SFCs. Here, we have used the python package {\it photutils} to perform aperture photometry. One of the important tasks for aperture photometry is the subtraction of the background. This is done by fitting apertures of the same size as the SFCs and the total galaxy in random source-free positions on the fits file. Then we calculate the background and subtract those from the actual counts. To calculate the total counts from a galaxy, we have used the semi-major axis from the 2MASS K-band image or the R$_{25}$ radius (Table \ref{magnitudes}). However, visual inspection was done before aperture photometry and whenever we find the UV emission more extended, we change the radius accordingly. These details are given in Table \ref{magnitudes}. This is to note that we have treated ESO 509-IG066 as a single system while performing the extinction calculation (Table \ref{magnitudes}, \ref{extinction}). However, as two galaxies (east and west) are still well separated, we have provided the total SFR of individual galaxies (Table \ref{total_sfr}). We have used the position as well as the size along with the orientation of the SFCs from {\it SExtractor} to perform aperture photometry (Table \ref{sfc_sfr}).

\subsubsection{Uncorrected Magnitudes and extinction}
The background-subtracted counts are converted into magnitude using the exposure time of the observations and the zero points, taken from \citep{tandon2020}. The uncorrected magnitudes are given in Table \ref{magnitudes}. Next, we have corrected the Milky Way extinction using E(B-V) $\times$ R$_V$, where E(B-V) is the reddening and R$_V$ is the extinction ratio. Here, E(B-V) for our sample galaxies are obtained from \citet{sf2011} which is available at IRSA page\footnote{https://irsa.ipac.caltech.edu/applications/DUST/}. We have taken R$_{FUV}$ and R$_{NUV}$ as 8.06 and 7.95 respectively \citep{bianchi2011}. In Table \ref{magnitudes}, we have provided the GALEX magnitudes (FUV, NUV) for comparison with the UVIT magnitudes as well as the Galactic extinction corrected UVIT magnitudes which are further used to calculate the color of the galaxies.

Galaxies have internal dust that also absorbs UV light. This is why the calculated flux is always lower than the actual value. Several different approaches are adopted to correct for this absorption. We have used UV spectral slope $\beta$ (f$_\lambda$ $\propto \lambda^{\beta}$) to calculate the color excess. The following equation is used to calculate $\beta$ from \citet{nordon2013}:
\begin{equation}
\beta = -\frac{m(\lambda_1) - m(\lambda_2)}{2.5 \log\left(\frac{\lambda_1}{\lambda_2}\right)} - 2
\end{equation}

Here $\lambda_1$, $\lambda_2$ are effective wavelength of the FUV and NUV filters and m$(\lambda_1)$, m$(\lambda_2)$ are the Milky way corrected magnitudes. More negative values of $\beta$ imply the least dust while more positive values imply the more dusty system. The calculated $\beta$ values are given in Table \ref{extinction}. The color excess is calculated following \citet{reddy2018}:
\begin{equation}
    \beta = -2.616 + 4.684E(B-V)
\end{equation}

Following \citep{calzetti2000}, the extinction A$_V$ is E$_s$(B-V)K$_\lambda$ where 
\begin{equation}
    K_\lambda = 2.659(-2.156 + 1.509/\lambda - 0.198/\lambda^2 + 0.011/\lambda^3) + R{^\prime}_V
\end{equation}
for 0.12 $\mu$m $\leq \lambda \leq 0.63$ $\mu$m and E$_s$(B-V) = (0.44 $\pm$ 0.03)E(B-V). The calculated values of A$_{FUV}$ A$_{NUV}$ are listed in Table \ref{extinction}. 

The above task is done for the total galaxy (radius: Table \ref{magnitudes}) and these single values of A$_{FUV}$ and A$_{NUV}$ of individual galaxies (Table \ref{extinction}) are used to correct the SFR of the total galaxy as well as the SFR surface density of the SFC. Calculating the extinction of individual SFCs is difficult, especially when the SFCs are not distributed evenly on the disk. Here, we have tried to estimate the maximum, and minimum values of average A$_V$ using annulus with increasing radius from center to the maximum radius where outermost SFCs are detected. These values are listed in Table \ref{extinction} for the reference. This is to note that the maximum and minimum A$_V$ values can increase and decrease the SFR of SFCs up to 2.33 times and 0.39 times respectively.

\subsubsection{Color maps and radial profiles}
We have FUV and NUV maps for eight galaxies while one has only FUV data. 
% The color maps can help us understand the stellar population as well as the merger effect on star formation i,e whether the gas inflow produces young stars at the center of the galaxies or it starts from the outskirts of the galaxies.
To create color maps, we first aligned the FUV and NUV images using {\it geomap} and {\it geotran} in IRAF. Then we used the background subtracted, integration-time-weighted images to create the NUV/FUV images. Next, we converted the NUV/FUV count ratio image to the magnitude scale using the zero point of the individual bands. 

\begin{figure*}
\centering
\includegraphics[width=18cm, trim = 0 7cm 0 0]{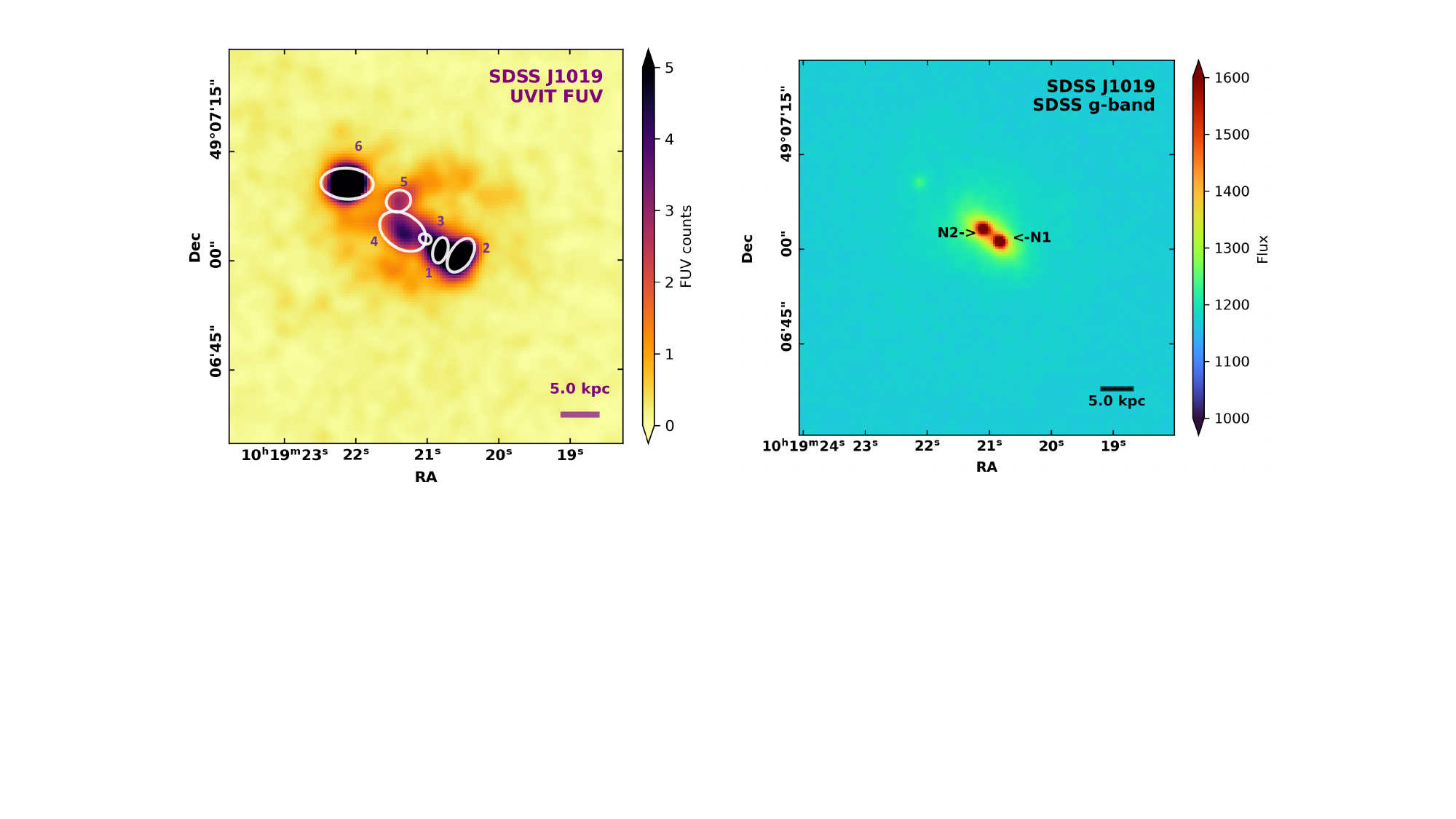}
\caption{\small Multi-band images of J1019. ({\it left:}) The UVIT-FUV image; ({\it right:}) The g-band optical image from SDSS. This system also hosts two AGNs. Though it is not an AGN-dominated system according to figure \ref{wise_color}, as it shows the highest FUV SFR, we masked the AGN at the centers of SFC 1 and 3 with a 3-pixel aperture to minimize the AGN contribution (Table \ref{total_sfr}, \ref{sfc_sfr}).}
\label{fig1019}
\end{figure*}

\begin{figure*}
\centering
\includegraphics[width=18cm, trim = 0 0 0 0]{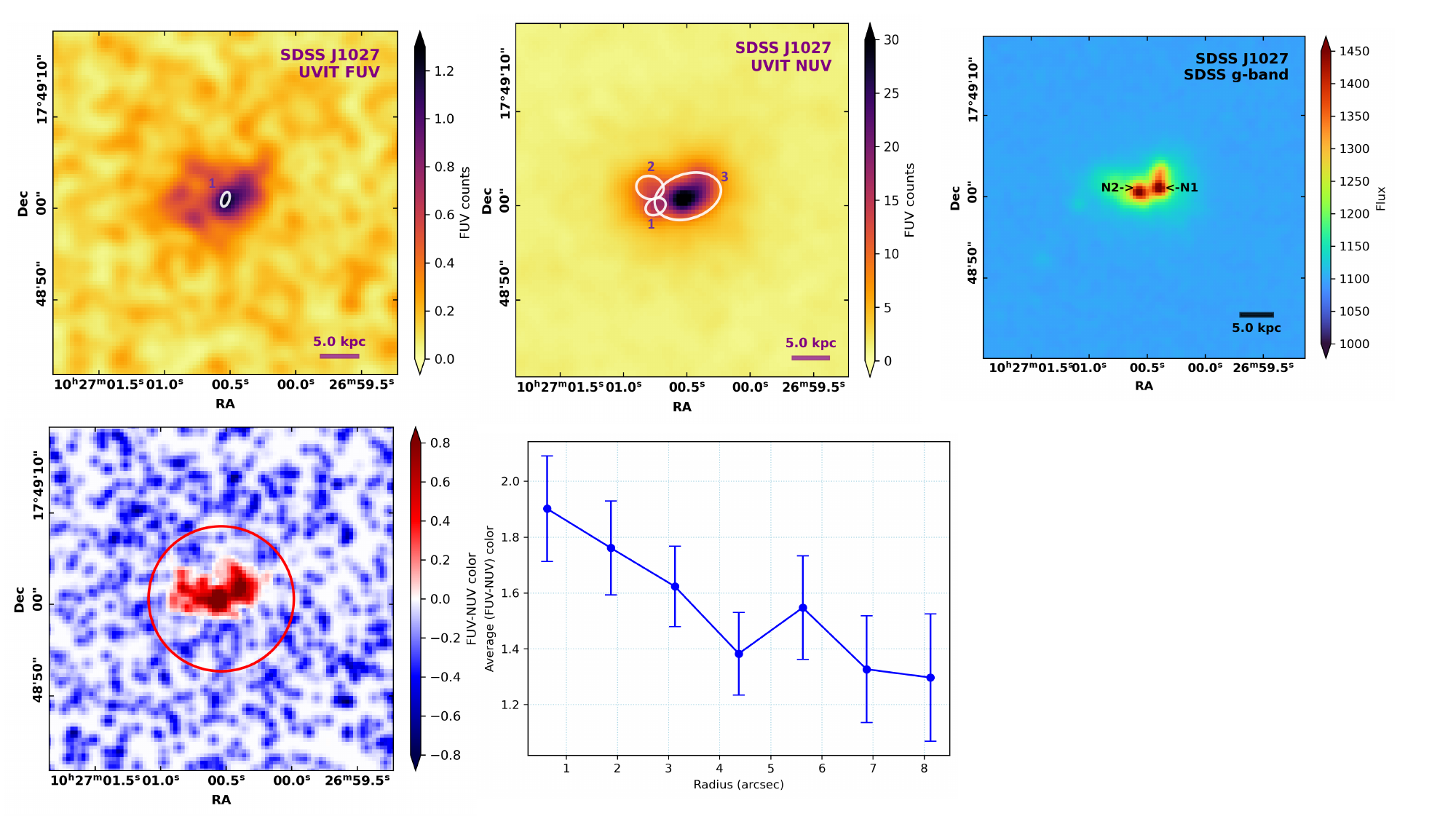}
\caption{\small Multi-band images of SDSS J1027.  Same as figure \ref{fig306}.}
\label{fig1027}
\end{figure*}

To understand the color profile of the galaxies more quantitatively, we have calculated the average color in consecutive annuli increasing outwards radially. We started from the galaxy centers and then used circular annuli from {\it photutlis} and did aperture photometry for individual annuli in NUV and FUV bands. The annulus radius is kept at 3 pixels and extends up to the galactic emission (red circles in lower left images: Figure \ref{fig306} to \ref{fig1027}). Then we converted the background subtracted counts to the magnitudes as above and calculated the color by subtracting the FUV - NUV magnitude of the individual annuli. Finally, to examine the color profile, we plotted the average color with the aperture radius (lower middle: Figure \ref{fig306} to \ref{fig1027}). 

\subsection{Star-formation rate and stellar mass estimation}
The SFR and stellar mass provide crucial information about any galaxy. Hence, proper estimation of star-formation rate (SFR) is one of the most essential parts of understanding galaxies and their evolution. While the global SFR of a galaxy provides information about the global properties and is related to galaxy evolution, the local SFR within the galaxies helps us to understand the spatial variations, the SF trigger mechanisms as well as any feedback process. There are many calibrators to calculate SFR, but the UV photons emitted by the young stars are often taken as a good indicator of recent SFR. However, as mentioned above, the extinction correction is needed. We have derived UV extinction corrected SFR using the following formulas \citep{iglesias2006, cortese2008} assuming a Salpeter initial mass function (IMF) from 0.1 to 100 M$\odot$ and solar metallicity:

\begin{equation} 
\mathrm{SFR_{FUV}(M_\odot yr^{-1}) = {\frac{L_{FUV}(erg\  s^{-1})}{3.83\times 10^{33}}}\times 10^{-9.51}}
\end{equation}

\begin{equation} 
\mathrm{SFR_{NUV}(M_\odot yr^{-1}) = {\frac{L_{NUV}(erg\ s^{-1})}{3.83\times 10^{33}}}\times 10^{-9.33}}
\end{equation}

The extinction-corrected total SFR as well as the SFR surface density for individual SFCs are given in Table \ref{total_sfr} and Table~\ref{sfc_sfr} respectively. In the next section, while discussing the SFR, we also reference the SFR from available literature for comparison. 

Although some of the sample galaxies have stellar masses available in the literature, they have been obtained using different survey data as well as different SED fitting methods. So to avoid this inconsistency, we have calculated stellar masses using WISE W1 - W2 color and W1 luminosity following \citet{cluver2014}:

\begin{equation}
\mathrm{log_{10} M_{Stellar} / L_{w1} = -2.54(W_{3.4 \mu m} - W_{4.6 \mu m}) - 0.17}
\end{equation}
with 
\begin{equation}
     L_{\text{W1}} (L_\odot) = 10^{-0.4(\text{M} - \text{M}_{\text{SUN}})}
\end{equation}

where M is the absolute W1 magnitude of the source in W1 and M$_{\text{SUN}}$ = 3.24. The stellar masses are listed in table \ref{stellar_mass}.
We have calculated the specific star-formation rate (sSFR = SFR/M$_\star$) of our galaxies (Table \ref{stellar_mass}) using the total extinction corrected FUV SFR (Table \ref{total_sfr}) and stellar mass.

\begin{figure*}
\includegraphics[width=18cm, trim= 0 8cm 0 0]{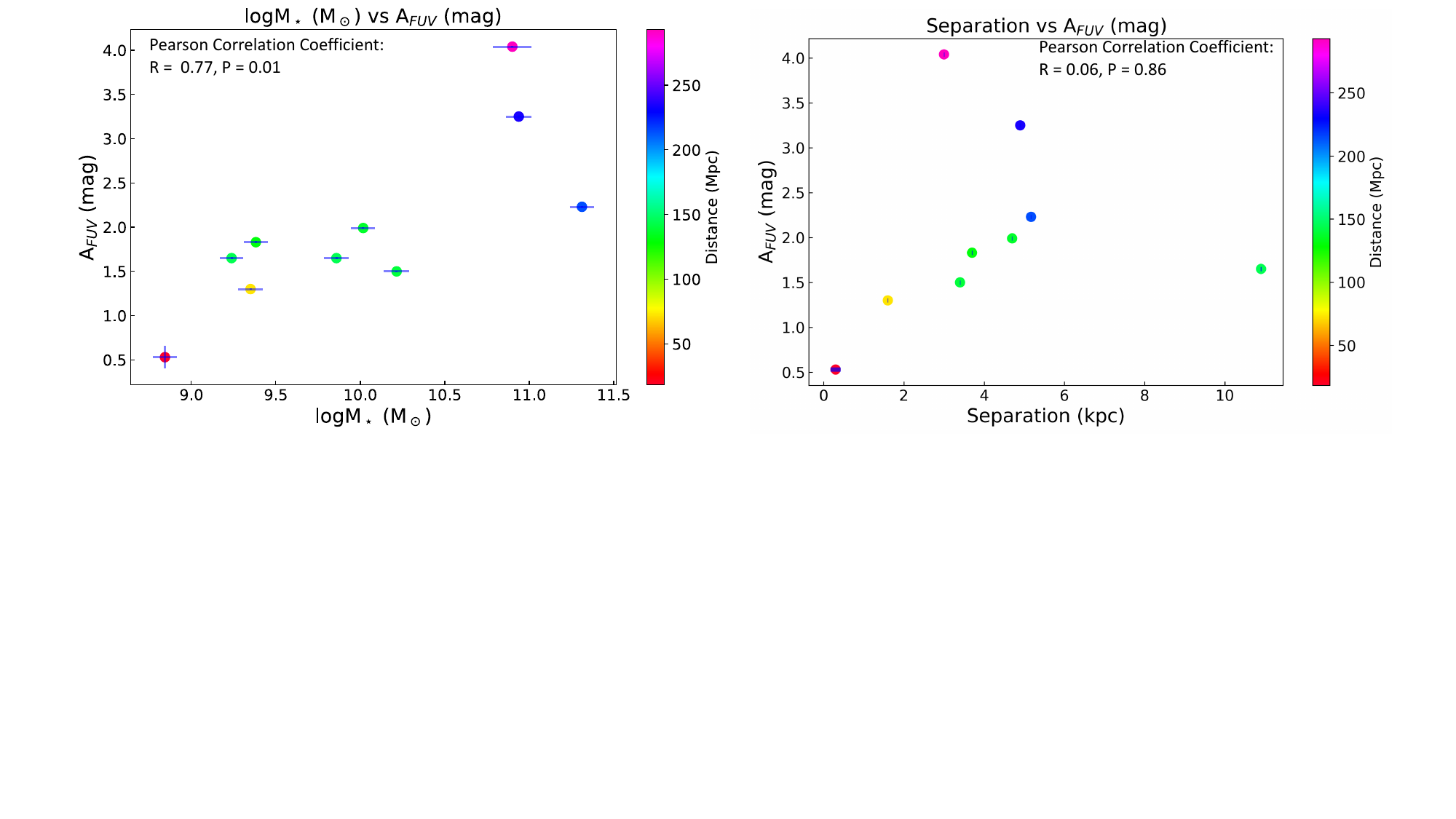}
\caption{\small The correlation between the extinction and stellar mass ({\it left}), separation of the nuclei ({\it rught}). While it shows a good correlation with the stellar mass, there is no correlation with the nuclear separations. The correlation coefficients are given in the upper corners.}
\label{correlation_Afuv}
\end{figure*}

\begin{figure*}
\includegraphics[width=17cm, trim= 0 7cm 0 0]{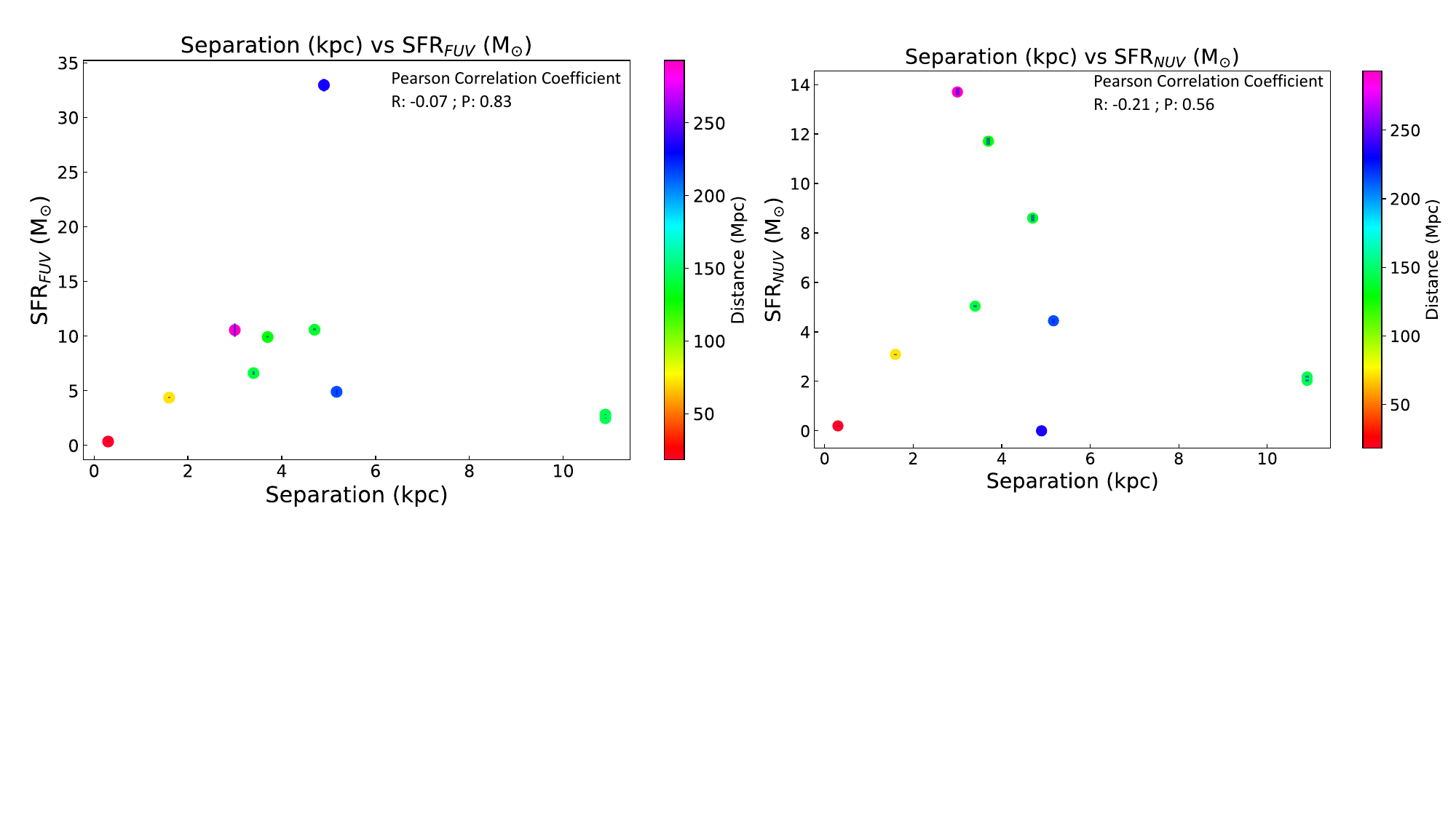}
\caption{\small The correlation between the SFR$_{FUV}$ ({\it left}), SFR$_{NUV}$ ({\it rught}) with the nuclear separation. It shows no correlation.}
\label{correlation_sfr}
\end{figure*}

\begin{figure}
\includegraphics[width=0.95\columnwidth, trim= 10cm 1.5cm 9cm 0]{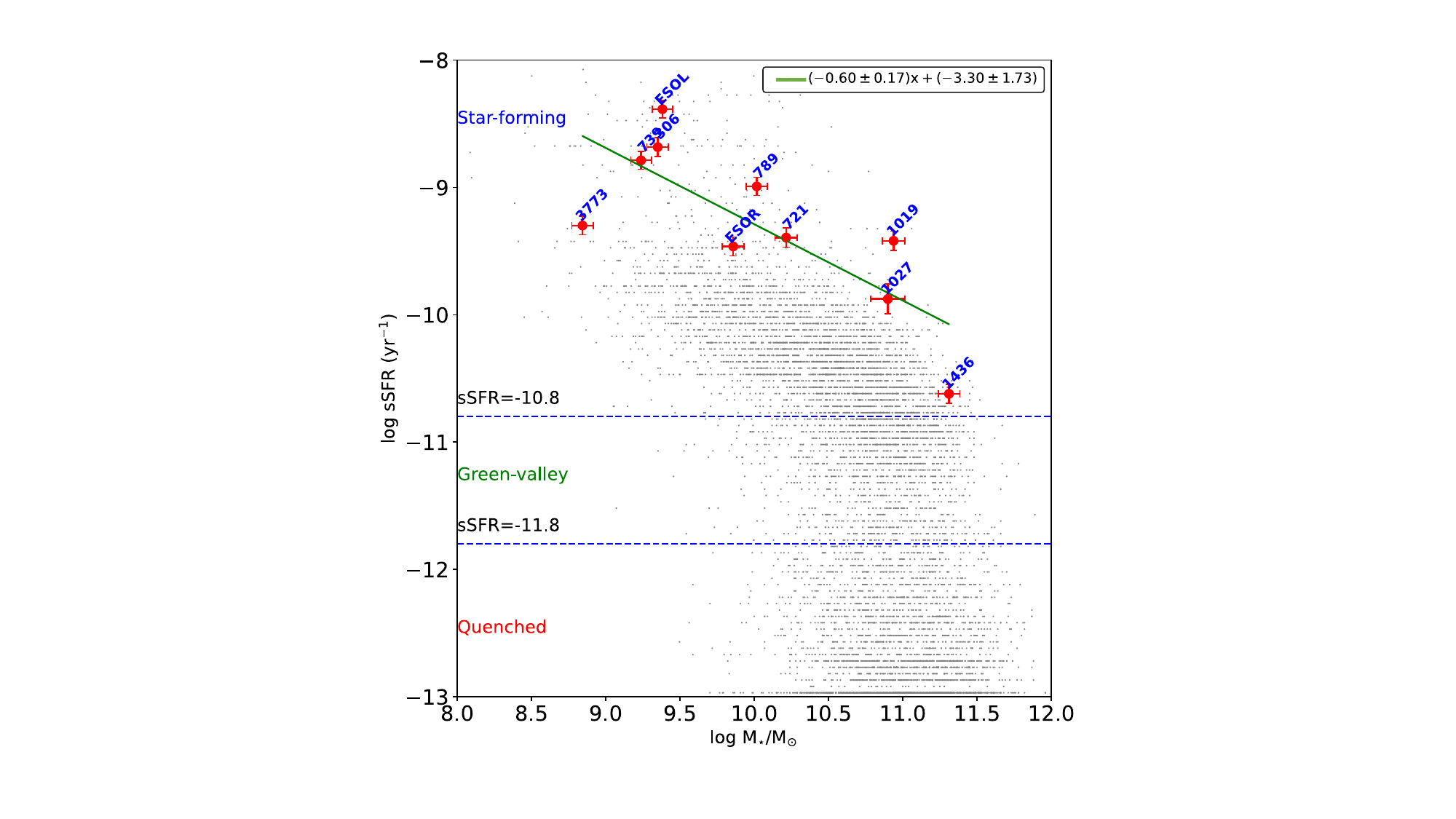}
\caption{\small The logarithmic plot of specific SFR (sSFR) versus stellar mass (M$_\star$) (Table \ref{stellar_mass}). Our sample galaxies are in red dots. The star-forming region and green valley are divided at sSFR $=-10.8$ while the quenched region of the plot is divided at sSFR $=-11.8$. All of our sample galaxies are in the star-forming regions. The gray points in the background are galaxy data taken from \citet{Bait17}.}
\label{fig_MS}
\end{figure}

\begin{figure*}
\includegraphics[width=18cm, trim= 0 8cm 0 0]{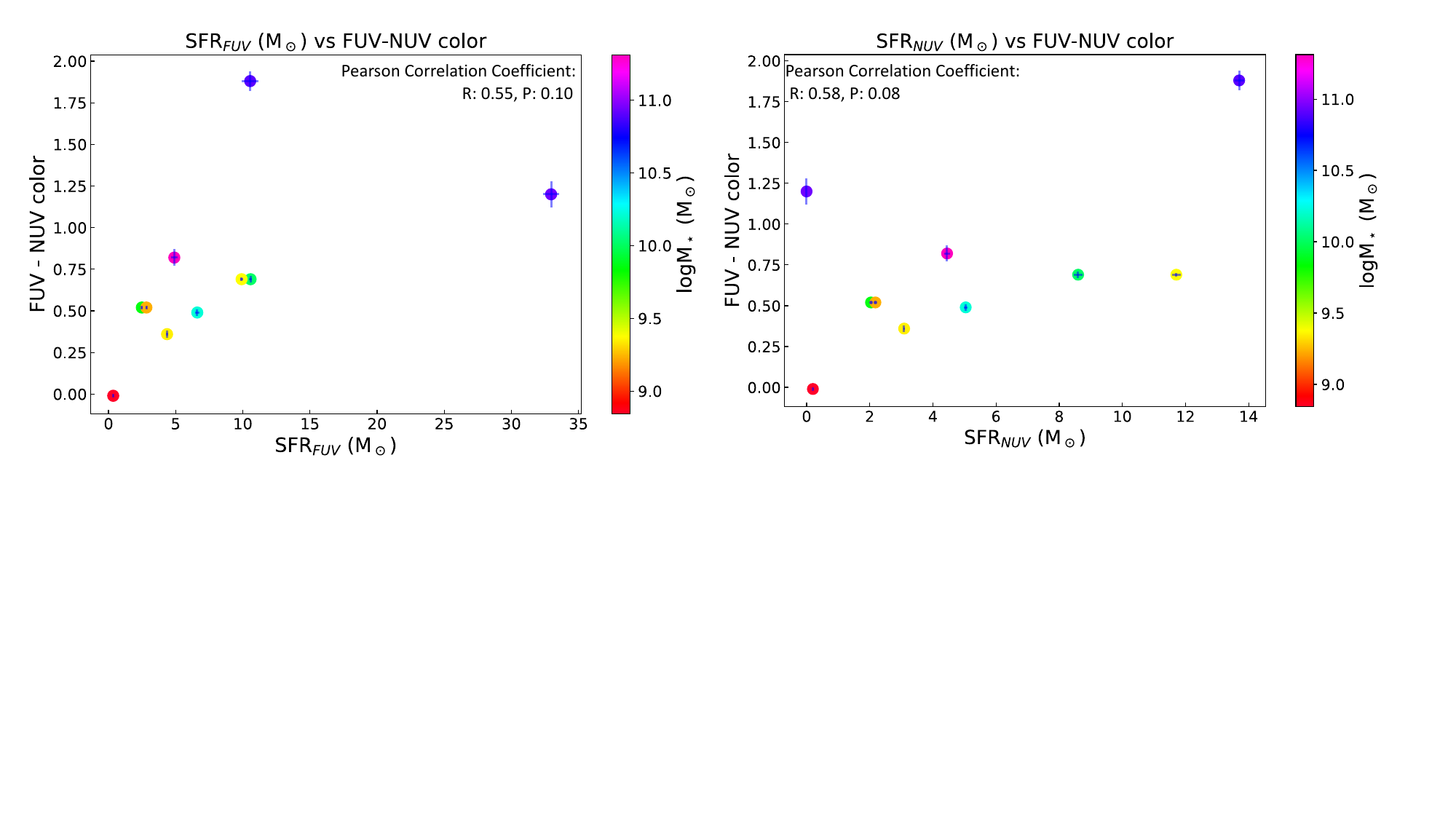}
\caption{\small The correlations between FUV-NUV color and SFR$_{FUV}$ ({\it left}), SFR$_{NUV}$ ({\it right}). These show a mild correlation.}
\label{correlation_color}
\end{figure*}

\section{Results and discussion}
%\subsection{Sample Galaxies with UV Images}
\subsection{Extinction corrected global SFR and sSFR} \label{global_SFR}
Star formation in galaxies produces UV photons in bulk but a significant fraction of the light is absorbed by interstellar dust. \citet{buat2009} suggested that the internal extinction in galaxies can be sometimes quite large and should be taken care of while using UV emission to estimate SFRs. We have attempted to calculate the dust attenuation in our sample galaxies using the UV continuum slope ($\beta$) which is assumed to be the deviation from the inherent recent star-formation \citep{meurer1995, buat2009, hao2011, pannella2015}. In a recent study with the UVIT, \citet{pandey2021} have found $\beta$ to range from $-2.72$ to $-0.60$ for six star-forming galaxies in the Bootes Void ($z\sim0.043 - 0.57$). 

\citet{gold2002} studied bright IR galaxies and calculated $\beta$ values at different aperture radii. While the $\beta$ range from $-1.42$ to 4.41 in the central 0.5 Kpc, the average values over the total galaxy are $\beta=-1.35$ to $-0.09$. This shows that dust attenuation varies from the center of galaxies to the outskirts. Similarly, \citet{yuan2012} studied a sample of galaxy pairs to understand the dust properties during mergers and compared it with a control sample of isolated galaxies. They found that while the $\beta$ value ranges from $\sim-1.5$ to $\sim1$ in isolated galaxies, the galaxy pairs have more scatter in $\beta$, with values ranging from $\sim-3$ to $\sim3$. 

The calculated $\beta$ of our sample galaxies falls in the range of the above-mentioned studies. Our calculated $\beta$ parameters range from $-2.03\pm0.13$ to $1.53\pm0.01$ (Table \ref{extinction}). The more negative $\beta$ implies less dust obscuration while values close to zero or positive values imply more dust obscuration. As dual nuclei are the expected outcome of galaxy mergers, dust is expected in these galaxies. The calculated A$_{FUV}$ and A$_{NUV}$ ranges from $0.53\pm0.13$ to $4.04\pm0.13$ and $0.44\pm0.10$ to $3.10\pm0.01$ for our sample (Table~\ref{extinction}). These values are calculated for the total galaxy with the aperture given in Table~\ref{magnitudes}.

We have also explored the correlation between A$_{FUV}$ vs stellar mass (M$_\star$) and the separation of the nuclei (Figure \ref{correlation_Afuv}). This is because a smaller separation of galaxy pairs may indicate a later stage in galaxy mergers. Hence, both the dust mass, as well as dust distribution, can be different in the later stages of galaxy mergers compared to the early stages. We find that stellar mass and A$_{FUV}$ are correlated with a correlation coefficient value of R = 0.77, P = 0.01. However, there is no significant correlation between A$_{FUV}$ and separation with R = 0.06 and P = 0.86. \citet{yuan2012} found a range of A$_{FUV}$ = 0.66 to 5.26 for their galaxy pair sample. They have also checked the dependencies of A$_{FUV}$ on the separation of the pairs along with the stellar mass. Their data show that A$_{FUV}$ increases with stellar mass but it did not show any dependency on separation.

We have corrected the observed FUV and NUV fluxes for the Milky Way extinction as well as galaxy internal extinction for the sample. As some of our sample galaxies have very high extinction values, the SFRs increase by a factor of $\sim$2 to $\sim$48 times. One such example is the study of post-merger galaxy NGC 7252 by \citet{george2018}. They have found that the SFR derived from FUV emission from the main disk is 0.66$\pm$0.01 M$\odot$ yr$^{-1}$ whereas other indicators show much higher values \citep[SFR IR/1.4~GHz/H$\alpha$: 8.1/(6.3$\pm$0.2)/(5.6$\pm$1.1) M$\odot$ yr$^{-1}$][]{schweizer2013}. Furthermore, their estimation of A$_{FUV}$ turned out to be 2.33 mag which can attenuate the actual UV light by 8 times. 

The extinction corrected SFR for both the bands are given in table \ref{total_sfr}. SDSS J1019 showed the highest FUV SFR. Here, we have re-estimated the SFR after masking the nuclei.
The final SFR ranges from $\sim$0.35$\pm$0.01 to 32.96$\pm$0.62 M$\odot$ yr$^{-1}$. 
%with a median value of 6.12 M$\odot$ yr$^{-1}$. 
The NUV SFR ranges from 0.203$\pm$0.001 to 13.70$\pm$0.15 M$\odot$ yr$^{-1}$. 
%with the median value of 4.45 M$\odot$ yr$^{-1}$. 
The high SFR may be an outcome of the star formation associated with the galaxy mergers. This is just to mention that the sample galaxies are also UV-rich according to our selection criteria (section \ref{sample_selection}). Further, we have tried to compare the SFR with the available literature. However, it should be noted that there are several methods to estimate the SFR which can produce quite different values depending on the inputs. 

We have found an extinction corrected SFR$_{FUV}$ = 9.91$\pm$0.10 M$_\odot$ yr$^{-1}$ for MRK 739 where the emission from both nuclei ($\sim$ 1.2$\arcsec$) are masked. The uncorrected SFR is 1.55$\pm$0.01 M$_\odot$ yr$^{-1}$. The SFR$_{FIR}$ is reported as 6.9 M$_\odot$ yr$^{-1}$ in \citet{tubin2021}. The triple AGN candidate SDSS J1027 in our sample has a SFR$_{FUV}$ = 10.55$\pm$0.62 M$_\odot$ yr$^{-1}$ whereas \citet{foord2021b} reported SFR$_{IR}$ as 18.2$\pm$1.3 M$_\odot$ yr$^{-1}$. MRK 789 is a starburst merger galaxy, and we have found the SFR$_{FUV}$ = 10.58$\pm$0.21 M$_\odot$ yr$^{-1}$, whereas the SED produced SFR is 5.99$\pm$1.12 M$_\odot$ yr$^{-1}$ \citep{salim2016}. The overall analysis shows that the dual nuclei sample galaxies are dusty systems with high SFRs which may be the outcome of mergers. 

Galaxy SFRs are found to increase with decreasing galaxy separation in some of the large statistical samples \citep{ellison2008}. However, it depends on several factors such as galaxy masses, their mass ratios, prograde/retrograde orbits, and gas content. 
%\citet{stemo2020} studied a sample of 220 dual and offset AGN candidates where SFR$_{MEDIAN}$ ranges from 0.04 to 84.78 M$_\odot$ yr$^{-1}$ and found no correlation between the SFR and bulge separation in their final results.
We checked whether the SFRs have any correlation with the separation of the nuclei. It shows that there is no correlation as the Pearson correlation coefficient is R = -0.07 and P = 0.83 for SFR$_{FUV}$; also R = -0.21 and P = 0.56 for SFR$_{NUV}$ (Figure \ref{correlation_sfr}). However, it should be noted that our sample number is very small and inhomogeneous in terms of redshift or separation. Hence, this result is limited and hard to compare with the results from larger unbiased samples.

The range of the calculated stellar masses is (6.98$\pm$1.27)$\times$10$^8$ to (2.05$\pm$0.37)$\times$10$^{11}$ M$_\odot$ (Table \ref{stellar_mass}).
%While some of these values match with the other literature \citep{salim2016}, some of our estimated M$_\star$ is lower than the estimation from other literature (e.g. MRK 739: \citet{tubin2021}).
The stellar mass (M$_\star$) and specific SFR (sSFR) are correlated for the star-forming galaxies and this is known as the star-forming main sequence. The sSFR-M$_\star$ is one of the most important parameters to estimate the current level of star formation in a galaxy for the available stellar material. It also indicates whether a galaxy is going through extensive star formation or is in a quenched phase. Our calculated log(sSFR) ranges from -10.26$\pm$0.07 to -8.38$\pm$0.07 yr$^{-1}$. All of the sample galaxies fall in the star-forming region (Figure \ref{fig_MS}) which means that they are still actively forming stars. This is one of the expected results of our study of star formation in dual nuclei galaxies. We have plotted M$_\star$-sSFR (Figure \ref{fig_MS}) with a control sample from \citet{Bait17} which studied 6000 galaxies in the local Universe to understand the dependence of star formation on the morphological types.  

We have fitted the data points in Figure \ref{fig_MS} and found a trend of sSFR decreasing with increasing stellar mass. Next, we checked the Pearson Correlation Coefficient which turned out to be R $=-0.775$, P = 0.008 signifying an anti-correlation. We fitted the data points with log(sSFR) yr$^{-1}$ = ($-0.60\pm0.17$) logM$_\star$ M$_\odot$ + ($-3.30\pm1.73$). \cite{yuan2012} have found a similar trend where x, y $= -0.30, -7.28$ for their spiral pairs and x, y $= -0.53, -4.94$ for the control sample.

\citet{tubin2021} studied dual AGN MRK 739 with high-resolution optical spectroscopy. Using SFR$_{H\alpha}$ emission, they found that the western nucleus forms stars at a rate of 5 M$_\odot$ yr$^{-1}$ while the eastern nucleus is quenched. The well-known dual AGN-host galaxy and merger remnant NGC 6240 shows an SFR of 100 M$_\odot$ yr$^{-1}$ \citep{muller2018}. Galaxy mergers or post-merger systems are found to show intense ongoing SFR. We have found similar results for our sample galaxies.

\subsection{SFCs properties} \label{SFC}
We utilized the higher-resolution UVIT images to probe star-forming regions in our sample galaxies. In the past few years, different studies have analyzed star-forming clumps in nearby galaxies of different morphologies using UVIT observations to understand the SFC properties and distributions. \citet{rahna2018} observed the barred spiral galaxy NGC 2336 with different filters of UVIT which is at a distance of 32.2 Mpc and has a size of 7.1$^\prime \times 3.9^\prime$. They detected 78 individual knots in NUV and 57 knots in FUV (which they call star-forming knots) with mean sizes of 485 pc and 408 pc in the FUV and NUV bands. \citet{mondal2018} studied the nearby irregular dwarf galaxy WLM (distance: 995 kpc) with UVIT and detected several possible young stellar associations with $4-50$~pc sizes. \citet{yadav2021} studied three nearby galaxies at distances varying between $\sim6-7.5$~Mpc, with R$_{25}$ ranging from 4.89$^\prime$ to 12$^\prime$ and a few hundreds of SFCs were detected with sizes ranging from a few parsecs to kpc and SFR density ranging from $\sim$ 10$^{-3}$ to 10$^{-1}$ M$_\odot$ yr$^{-1}$ kpc$^{-2}$. 

Most of these studies investigated the SFCs using UVIT in nearby large galaxies with very few galaxies around or beyond the distance of 70 Mpc. One such distant UVIT galaxy study is by \citet{rakhi2023} where 56 knots with sizes $\sim$1 Kpc$^2$ to $\sim$35 Kpc$^2$ were detected in NGC 5291, which is at a distance of 62 Mpc. In our sample, while one galaxy, NGC~3773 is at a distance of $\sim18$ Mpc it has a size of 40$\arcsec$, and {\it SExtractor} has detected only one SFC in the disk of the galaxy (Figure \ref{fig3773}). The remaining eight galaxies of our sample are situated at distances ranging from 73 Mpc to 293 Mpc and have a radius of $\sim$ 20$\arcsec$ - 50$\arcsec$.  We have detected 1 to 14 SFCs in FUV images and 1 to 19 SFCs in NUV images of our galaxies. There is no SFC detected only on the FUV image of SDSS J1436. The area size varies from 0.18 Kpc$^2$ to 42 Kpc$^2$. Most of the SFCs are situated in the disk. The calculated extinction corrected surface density ($\Sigma_{SFR}$) ranges from $\sim$ 0.0064$\pm$0.0004 - 0.9069$\pm$0.0079 M$_\odot$ yr$^{-1}$ kpc$^{-2}$ (Table \ref{sfc_sfr}) which overlaps with $\Sigma_{SFR}$ values of nearby spiral galaxies like NGC 628 \citep{yadav2021}.

\begin{figure}
\includegraphics[width=\columnwidth, trim= 0cm 0cm 0cm 0]{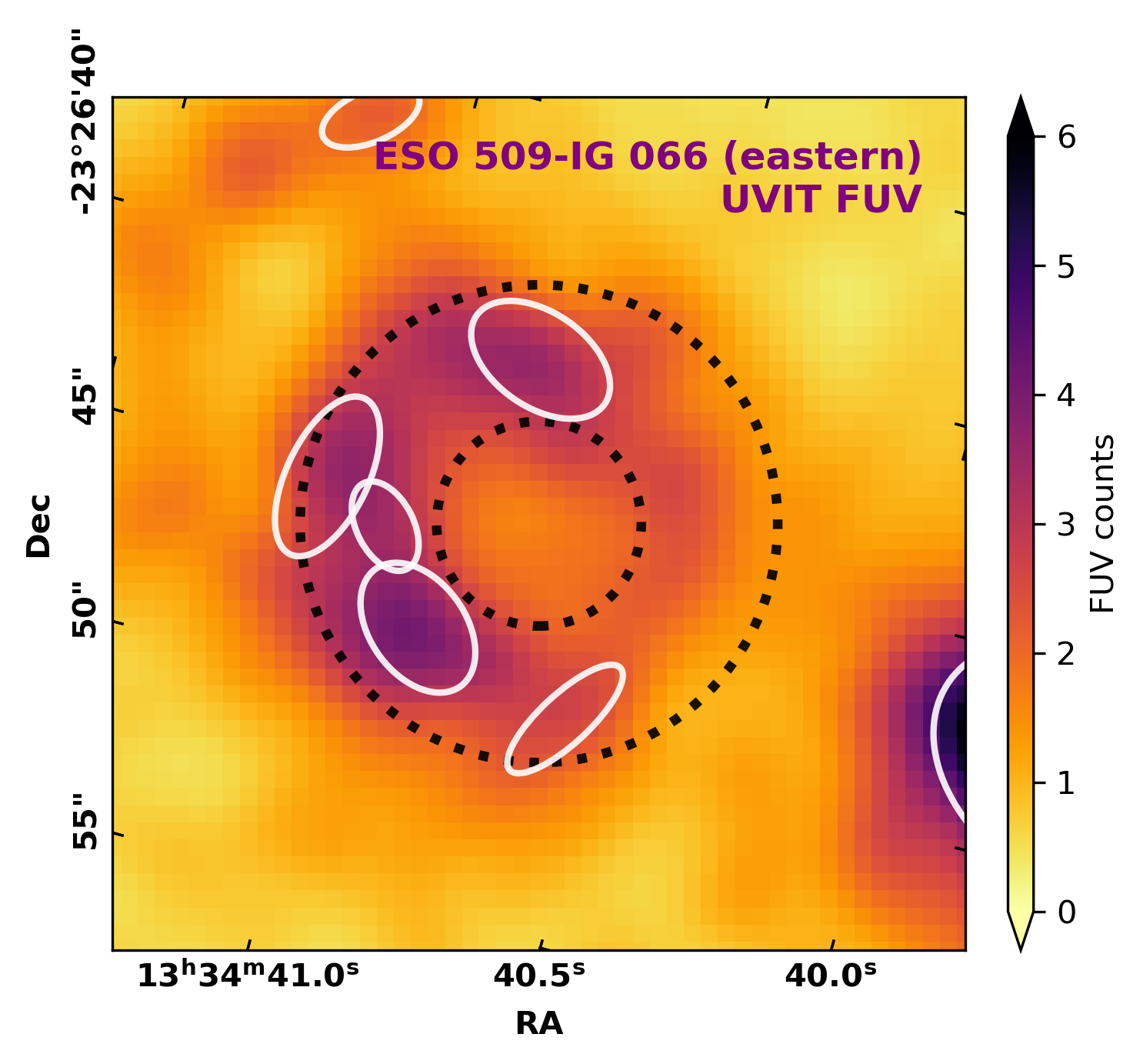}
\caption{\small The FUV image of the eastern galaxy of ESO509-IG066 merger system. The white elliptical apertures are the SFCs as shown in figure \ref{figESO}. The black dotted ring is an artistic view: the star formation is happening in a ring form which can be connected with central AGN which shows conical outflows. AGN may blow out the gas from the center which gets compressed and cools down to form new stars in the ring.}
\label{eso_ring}
\end{figure}

\subsection{FUV- NUV color} \label{color}
The FUV$-$NUV color maps can help us understand the stellar population in galaxies. Several UVIT studies of nearby galaxies have used the FUV$-$NUV color to calculate the age of the resolved clumps. However, as we have sample galaxies that are small in size and are at large distances when we matched the resolution and created the color maps, most of the clumps disappeared. Here, we have analyzed the color maps with the radial profile of FUV$-$NUV color (Figure \ref{fig306}$-$\ref{fig1027}).  \citet{gil2007} studied $\sim$ 1136 galaxies in the GALEX field and examined their FUV$-$NUV color. With an FUV$-$NUV color range of $\sim-0.20$ to 3.0, they found a peak around (FUV$-$NUV) $\sim0.4$ mag as well as a cut-off at (FUV$-$NUV)= 0.9 mag which separates elliptical/lenticular galaxies from spirals galaxies. 

The total FUV-NUV color is calculated using Galactic extinction corrected FUV and NUV magnitudes as given in table \ref{magnitudes} and it ranges from $\sim$ 0 to 1.88 mag which is similar to that seen in other nearby galaxies \citep{boselli2005a, boissier2018}. The color maps (lower left: Figure \ref{fig306} to \ref{fig1027}) show that all of our sample galaxies except one show a redder color in the center i.e. the FUV-NUV value is higher in the center while the outskirts are relatively bluer (lower FUV-NUV value). NGC 3773 is the only galaxy that shows a bluer color at the center and has a redder ring around it (Figure \ref{fig3773}: lower left and middle). The variation of color in our sample galaxies, from inward to outwards implies the following (i) the central disk has more old stars than young stars, (ii) these galaxies are in a merger phase where the star formation is happening predominantly in the outskirts, (iii) the dust attenuation varies throughout the galaxy. Our galaxies have high extinction and there is a spatial variation in A$_V$ of individual galaxies as seen from maximum and minimum A$_V$ values (Table \ref{extinction}). However, with the present data, we cannot rule out other possibilities. 

We checked whether any correlation exists between color with stellar mass and FUV, NUV SFRs. The correlation between color and the stellar mass is the same as the correlation between the A$_V$ vs stellar mass (Figure \ref{correlation_Afuv}). The SFR$_{FUV}$, SFR$_{NUV}$ are mildly correlated with color with  R = 0.55, P = 0.10 and  R = 0.58, P = 0.08 respectively (Figure \ref{correlation_color}).

We did notice that two of the AGN-hosted galaxies (viz., SDSS J1019, SDSS J1027) show extreme color as well as extinction. However, these are the most distant as well as massive systems. Also, the other two AGN-hosted galaxies (MRK 739 and ESO509-IG066) show similar values to the star-forming galaxies. As our sample number is very small, we can not draw any strong conclusions on the active and non-active galaxies. We will explore it in the future.

The interacting galaxy, NGC4438 \citep{boselli2005} shows a 20 kpc long tail which is $\sim$ 20 kpc away from the nucleus. \citet{boselli2005} found recent star formation in tails and arms. However, most of the regions in the disk of the galaxy are dominated by old populations.
\citet{smith2010} studied the star formation morphology and stellar population of 42 interacting systems in the nearby Universe. They found that tidal tails in the outskirts are bluer than the parent disk which can be due to enhanced star formation due to interaction \citep{smith2010}. The calculated FUV-NUV color of our sample galaxies overlaps with the study by \citep[e.g.][]{smith2010, gil2007} as well as follows the same trend i.e. redder disk and bluer outskirts. However, we could not perform the age analysis of spatially distributed SFCs \citep{mondal2018, ujjwal2022}. Hence, we can not rule out whether different stellar populations and/or dust attenuation are causing the color variation.

\subsection{Looking for AGN/stellar Feedback} \label{feedback}
With three times better resolution than GALEX, UVIT shows resolved star-forming knots in a number of nearby galaxies. In \citet{rubinur2021}, one of our UVIT sample galaxies MRK 212 showed a pair of star-forming clumps near the central AGN. This coincided with the radio emission having a flat spectral index. This could be a signature of AGN-jet-induced SF. In recent work, \citet{nandi2023} detected three star-forming regions close to the nuclei of the AGN-hosted dwarf galaxy NGC 4395 which can be due to the AGN feedback effect. \citet{joseph2022} detected star-forming sources in the direction of the radio jet of Centaurus A suggesting AGN feedback. \citet{koshy2018} have found signatures of the possible suppression of SF in NGC 7252 and explained it with possible AGN feedback. NGC 7252 shows red nuclei (Avg color: $\sim$1.0), followed by the inner bluer ring (Avg color: $\sim$0.45) and then the redder outer ring (Avg color: $\sim$1.4). This was explained as an AGN blowing out gas from the nuclear region which forms new stars in the inner blue ring \citep{koshy2018}. Here it should be noted that ring-like structures are present in many galaxies which are mainly the product of galaxy interactions, or stellar bars \citep{combes1985, buta1996}.

We considered investigating feedback in our sample galaxies, although it is difficult as most of them are at a larger distance than the galaxies discussed above. Only in NGC 3773 which is $\sim$18 Mpc away, we found a redder stellar ring around a blue nuclear emission. In NGC 3773, the radial profile shows a value close to zero (redder) at the center and then it goes down to $-0.2$ (bluer) followed by rising to 0.6 (redder). Although we can not see any variation in the blue nuclei itself (and it is hard to explain 0 to $-0.2$) and there is a small difference in color to draw any conclusions (unlike in NGC 7252), the blue nuclei and red ring may indicate that the center has more recent SF considering the calculated extinction is very small in this galaxy. As previously mentioned, \citet{linden2020} state that their sample including NGC 3773 is producing stars in the central 500 pc regions. A nuclear supernova or star cluster may be ionizing the gas and hence producing new stars that show a blue color in the central region. \citet{gao2023} explored the star formation in the nearby dwarf galaxy merger system NGC 4809/4810 where they found that the star-forming knots surrounded by the supernova show the highest SFR. 

\citet{carlos2020} have detected a cone-shaped [O III] outflow in the eastern nuclei of ESO509-IG066. Our FUV image shows UV deficiency in the same nucleus and SFCs distributed in a tentative ring surrounding the nucleus (Figure \ref{eso_ring}). It is possible that outflowing materials is forming stars in the ring. The detailed study of MRK 739 by \citet{tubin2021} found that the eastern (left) nucleus is accreting as well as ionizing the northern regions and does not have much star-formation itself, whereas the western nucleus is not accreting but it falls in the star-forming main-sequence. They proposed that this system is consistent with an early stage of the galaxy collision, where the foreground galaxy (W) is a young star-forming spiral galaxy in an ongoing first passage with its background companion elliptical galaxy (E). In this scenario, the eastern nucleus is ionizing one of the north-western spiral arms,  similar to the “Hanny’s Voorwerps” phenomena, as seen in IC 2497 \citep{santori2016}. We have detected SFCs (Figure \ref{fig739}: SFC$_{FUV}$ id 3, 7 and SFC$_{NUV}$ id 5) in a similar position. This can be due to the AGN feedback process where the outflows associated with the eastern nucleus ionize the material and form stars in these SFCs.

\section{Summary \& Conclusions}
We have studied the UVIT images from AstroSat of 9 dual nuclei galaxies to understand star formation in them. Below is the summary with the main findings:
\begin{enumerate}
      \item Our sample galaxies are chosen with separations below or around 10 kpc where 8 galaxies have separation $<$ 10 kpc and one system has a separation of 11 kpc. From MIR color cut-offs, we checked whether AGN dominates the system and masked those nuclei.

      \item The UVIT magnitudes match the GALEX magnitudes. The calculated magnitudes are corrected for both internal and external (Milky Way) extinctions.  
      
      \item The external extinction is calculated from the UV continuum slope ($\beta$). The $\beta$ ranges from $-2.03\pm0.13$ to $1.53\pm0.01$. The total FUV extinction ranges from 0.53$\pm$0.13 to 4.04$\pm$0.01 and the NUV extinction ranges from 0.44$\pm$0.10 to 3.10$\pm$0.01 for the sample galaxies. 
      
     \item  The extinction-corrected total FUV SFR ranges from 0.350$\pm$0.002 to 32.96$\pm$0.62 M$_\odot$ yr$^{-1}$ and total NUV SFR ranges from 0.203$\pm$0.001 to 13.70$\pm 0.15$ M$_\odot$ yr$^{-1}$. The specific SFR put all of our sample galaxies in a star-forming region in the sSFR-M$\star$ plane and it decreases with the increase of stellar mass. 

     \item We have detected 1 to 14 SFCs in FUV images and 1 to 18 SFCs in NUV images. These SFCs are situated mostly in the disk and spiral arms. 
     
     \item The total FUV$-$NUV color of the galaxies ranges from $\sim0$ to 1.88. The color maps of most of the galaxies show redder emission in nuclear regions and bluer in the outskirts. The one dwarf galaxy, NGC 3773 shows bluer nuclei and an outer redder ring. The color variations in our sample galaxies can be either due to the dust variation or different stellar populations present which can be due to interactions. 

     \item The calculated extinction, SFR/sSFR, and color values of our sample galaxies fall in the range of galaxy merger systems from the literature. These quantities are correlated with stellar mass and no correlations are found between the SFR and the nuclear separation. However, it is difficult to draw any broad conclusion with a sample of nine galaxies. 

     \item We looked for any AGN or supernova-related feedback. The redder ring around the blue nuclei in the star-forming galaxy NGC 3773, the SFCs in a tentative ring-form around the east nuclei of dual AGN system ESO - IG066, or the SFCs on the spirals arm of MRK 739 indicate some signatures AGN/SNe feedback. However, we need to investigate further for confirmation with velocity information and/or finding the age of the populations, etc.\\

     % \item From the radio energy budget, we found that NGC3773 is a star-forming galaxy while MRK 789 is mostly hosting AGN. Other three galaxies (MRK 306, MRK 721, SDSS J1436) can have AGN but this needs to be investigated. The third nucleus in the triple AGN candidate J1027 is not detected in the high-resolution radio map. We could not resolve the left core of the ESO509-IG066 to investigate a newly born jet (present or not) that has a recent drop in accretion rate. \\
     
As our sample number is small and not homogeneous in terms of nuclear activity, stellar mass, or distance, we have avoided looking for any difference in active and non-active galaxies and this work is completely done to explore star formation in dual nuclei galaxies. In the future, with a larger and more homogeneous sample, we will explore the correlations concerning nuclei types in detail.
     
\end{enumerate}

\section*{Acknowledgements}
We thank the referee for their comments which have improved the manuscript significantly. We acknowledge IIA, NCRA and UIO for providing the computational facilities. RK, PK acknowledges the support of the Department of Atomic Energy, Government of India, under project 12-R\&D-TFR-5.02-0700. MD acknowledges the support of the Science and Engineering Research Board (SERB) Core research Grant CRG/2022/004531 for this research. This publication uses the data from the  AstroSat mission of the  Indian  Space  Research  Organisation  (ISRO),  archived at the  Indian  SpaceScience DataCentre (ISSDC) which is a result of collaboration between IIA, Bengaluru, IUCAA, Pune, TIFR, Mumbai, several centers of ISRO, and CSA.
The National Radio Astronomy Observatory is a facility of the National Science Foundation operated under a cooperative agreement by Associated Universities, Inc. This work has made use of the NASA ADS\footnote{https://ui.adsabs.harvard.edu/}. This research has made use of the  NASA/IPAC  Extragalactic  Database(NED), which is operated by the Jet Propulsion Laboratory, California Institute of Technology, under contract with the National Aeronautics and Space Administration. Funding for the Sloan Digital Sky Survey IV has been provided by the  Alfred  P.  Sloan  Foundation,  the  U.S. Department of Energy Office of Science, and the Participating Institutions. SDSS-IV acknowledges support and resources from the Center for  High-Performance  Computing at the University of Utah. The SDSS website is  www.sdss.org. This publication makes use of data products from the Two Micron All Sky Survey, which is a joint project of the University of Massachusetts and the Infrared Processing and Analysis Center/California Institute of Technology, funded by the National Aeronautics and Space Administration and the National Science Foundation.
RK thanks Prasanta Kumar Nayak for productive discussions.

\section*{Data Availability}
The raw data can be obtained from the following data archive: (i) https://astrobrowse.issdc.gov.in/astroarchive/archive/Home.jsp (ii) https://archive.nrao.edu/archive/advquery.jsp

\begin{table*}
\begin{center}

\caption{{Details of sample dual nuclei galaxies:} Columns: (2) Full name of the sample galaxies. Later on, shorter versions are used. (3, 4) RA (right ascension) and Dec (declination) in the J2000 epoch.(5, 6) Redshift (z) and distance. (7) Scale of the systems. (8, 9) Nature of the nuclei. (10, 11) Separations of the nuclei in arcsec and kpc scales. (12) References.}
\resizebox{\textwidth}{!}{
\begin{tabular}{ccccccccccccl}\hline
No &  Object & RA & DEC & Redshift & Distance & Scale & Nuclei 1 & Nuclei 2 & Separation &Separation & Reference\\

 & & (J2000) & (J2000) & & (Mpc) & (kpc/arcsec) & Type & Type & (arcsec) & (Kpc) & \\
 (1) & (2) & (3) & (4) & (5) & (6) & (7) & (8) & (9) & (10) & (11) & (12) \\  \hline 
1 & MRK 306 &  22h31m51.2s & +19$^{\circ}$41$\arcmin$9$\arcsec$ &  0.018696 &  72.9 &  0.341  & starburst &  unknown &  4.8 &1.6&\citet{mezcua2014}\\
2 & MRK 721 &  10h23m32.6s &  +10$\degr57\arcmin35\arcsec$ & 0.032092 & 140 & 0.636  & starburst & unknown & 5.4 &3.4 &\citet{mezcua2014}\\ 
3 & MRK 789 & 13h32m24.2s & +11$\degr06\arcmin23\arcsec$ & 0.031454 & 137  & 0.621  & AGN & unknown & 4.7 &2.9&\citet{mezcua2014} \\
4 & NGC 3773 & 11h38m12.9s & +12$\degr06\arcmin43\arcsec$ & 0.003276 &  18.3 & 0.088  & starburst & unknown & 3.2 &0.3 &\citet{mezcua2014}\\
5 & MRK 739 & 11h36m29.1s &+21$\degr35\arcmin46\arcsec$ & 0.029854 & 130  & 0.594  & AGN & AGN & 6.2 & 3.7&\citet{mezcua2014}\\
6 & ESO509-IG066 & 13h34m40.8s & -23$\degr26\arcmin45\arcsec$ & 0.033223 & 144  & 0.654  & AGN & AGN & 16 & 10.9&\citet{Guainazzi2005}\\
7 & SDSS J143648.10+182037.6 & 14h36m48.1s  & +18$\degr20\arcmin38\arcsec$ & 0.049421 & 214  &  0.940  & starburst & unknown & 5.5 & 5.2 &\citet{ge2012} \\
8 & SDSS J101920.83+490701.2 &  10h19m20.8s & +49$\degr07\arcmin01\arcsec$ & 0.054087 & 235  & 1.022  & AGN & AGN &4.6&4.9 & \citet{liu2011a}\\
9 & SDSS J102700.40+174901.0 & 10h27m00.4s & +17$\degr49\arcmin01\arcsec$ & 0.066619 & 293  &  1.245  & AGN & COMP & 2.4 &3.0 &\citet{liu2011a}\\ \hline
\label{sample}
\end{tabular}
}
\end{center}
\end{table*}

\begin{table*}
\caption{{UVIT observation details:} Columns: (1) Name of the objects. (2) UVIT proposal ID. MRK 739 and ESO509-IG066 have two observations. (3, 4): Name of the NUV filters and exposure time of the observations. (5, 6) Name of the FUV filters and exposure time of the observations.}
\begin{tabular}{ccccccccc}
\hline
Object & Proposal ID& NUV filter & NUV Exposure time  & FUV filter & FUV Exposure time\\ 

 & & & (secs) & & (secs) \\
 (1) & (2) & (3) & (4) & (5) & (6)  \\ \hline
 MRK 306 & A02\_165    & NUV-B13 &1228& Sapphire & 1446\\
 MRK 721 & A02\_165   & NUV-B13 &1819 & Sapphire  &1841 \\
 MRK 789 & A03\_091   &NUV-B13 &2137 & Sapphire &2137 \\
 MRK 739 & A03\_091 & NUV-B13 & 2258 & - & -\\
  & A07\_148  & -&-& CaF2 & 6500\\
 NGC 3773 & A03\_091   & NUV-B13 &2522 & Sapphire &2484 \\
 ESO509-IG066 &A03\_091  &NUV-B13 &2207 &Sapphire & 1833\\
  & A07\_148    &-&- & Sapphire  &13470\\
 SDSS J1436 & A03\_091     &NUV-B13 &2421 &Sapphire &2426 \\ 
 SDSS J1019 & A04\_085   &-&-& CaF2 &4451 \\
 SDSS J1027 & A04\_085  & Silica15 &  4986& CaF2 &4958 \\ \hline
\label{uvit_obs}  
\end{tabular}
\end{table*}

\begin{table*} 
\caption{{Magnitudes in FUV, NUV bands: Columns: (1) Name of the objects. (2) The radius of the apertures to calculate the UVIT magnitudes. These are taken as half of the major axis defined by 2MASS or R$_{25}$. However, visual inspections are done before the analysis, and apertures are adjusted from eye estimation depending on the UV emission where required$^{\star\star}$. (3, 4) The calculated magnitudes in the UVIT FUV and NUV bands. (5, 6) The GALEX FUV and NUV magnitudes. (7) The Galactic reddening. (8, 9) The Galactic extinction corrected UVIT magnitudes. As SDSS J1019 has only an FUV image from UVIT, we have corrected the GALEX magnitudes (FUV, NUV) and added them here. (10) The total color of the galaxies, calculated using columns 8 and 9.}}
\resizebox{\textwidth}{!}{
\begin{tabular}{lcccccccccc}
\hline
    & radius & \multicolumn{2}{c}{mag (UVIT)} &  \multicolumn{2}{c}{mag (GALEX)} &Galactic & \multicolumn{2}{c}{Corrected UVIT mag (Gal)} & Total color  \\
     \cline{3-4}   \cline{5-6} \cline{8-9} \cline{9-10}
Name  & (arcsec) & FUV & NUV &  FUV & NUV & E(B-V) & FUV$_{corr}$ & NUV$_{corr}$ & (FUV$_{corr}$-NUV$_{corr}$)  \\ 
 (1) & (2) & (3) & (4) & (5) & (6) & (7) & (8) & (9) & (10) \\ \hline
MRK 306 & 47&  16.45 $\pm$ 0.02 & 16.09 $\pm$ 0.01 & 16.51 $\pm$ 0.03 &  16.04 $\pm$ 0.02& 0.0488 & 16.06 $\pm$ 0.02 & 15.70 $\pm$ 0.01 & 0.36 $\pm$ 0.02  \\
MRK 721 & 21 &17.48 $\pm$ 0.03 & 16.99 $\pm$ 0.01 & 17.37 $\pm$ 0.05 & 16.81 $\pm$ 0.03 & 0.0232 & 17.29 $\pm$ 0.03 & 16.80 $\pm$ 0.01 & 0.49 $\pm$ 0.02\\
MRK 789 & 26 &17.38 $\pm$ 0.02 & 16.68 $\pm$ 0.01 & 17.40 $\pm$ 0.05 & 16.75 $\pm$ 0.02 & 0.0281 & 17.15 $\pm$ 0.02  &16.46 $\pm$ 0.01 & 0.69 $\pm$ 0.02\\
NGC 3773  &40  &15.21 $\pm$ 0.01 & 15.23 $\pm$ 0.01 & 15.23 $\pm$ 0.01 & 14.97 $\pm$ 0.01 & 0.0233 &15.03 $\pm$ 0.01 & 15.04 $\pm$ 0.01 & -0.01 $\pm$ 0.01 \\
MRK 739 &25$^{\star\star}$ & 17.77 $\pm$ 0.01 & 17.08 $\pm$ 0.01 & 17.46 $\pm$ 0.05 &  16.69 $\pm$ 0.02 & 0.0214  &17.60 $\pm$ 0.01 & 16.91 $\pm$ 0.01 & 0.69 $\pm$ 0.01\\
ESO509-IG066 & 20$^{\star\star}$ & 18.38  $\pm$ 0.01 & 17.84  $\pm$ 0.01 & 18.32 $\pm$ 0.10 & 17.73 $\pm$ 0.05& 0.0951 &  17.61 $\pm$ 0.01 & 17.09 $\pm$ 0.01 &  0.52 $\pm$ 0.01 \\
SDSS J1436 & 28 &  19.41  $\pm$ 0.05 & 18.59  $\pm$ 0.02 & 19.34 $\pm$ 0.13 & 18.57 $\pm$ 0.06& 0.0247& 19.21 $\pm$ 0.05  & 18.39 $\pm$ 0.02 & 0.82 $\pm$ 0.05 \\
SDSS J1019 & 14$^{\star\star}$ & 18.60  $\pm$ 0.02 & - & 18.59 $\pm$ 0.08    &  17.38 $\pm$ 0.03 & 0.0090 & 18.51 $\pm$ 0.08  & 17.31 $\pm$ 0.03 & 1.20 $\pm$ 0.08 \\
SDSS J1027 &19  & 21.19  $\pm$ 0.06 & 19.31  $\pm$ 0.01 & 21.07 $\pm$ 0.35  &  19.61 $\pm$ 0.13 &0.0208 &21.03 $\pm$ 0.06 & 19.15  $\pm$ 0.01 & 1.88 $\pm$ 0.06 \\ \hline
\label{magnitudes}  
\end{tabular}}
\end{table*}

\begin{table*}
\caption{{Internal extinction calculation:} Columns: (1) Name of the objects. (2) The UV spectral slop ($\beta$). (3) The calculated reddening (E(B-V)). (4, 5) The total extinction for FUV and NUV bands. This is done with the radius quoted in column 2 of table \ref{magnitudes}. The fluxes are corrected for these extinction values before calculating the SFR (Table \ref{total_sfr}) and $\Sigma_{SFR}$ (Table \ref{sfc_sfr}). (6$-$11): The maximum, minimum, and mean values of extinction in the FUV and NUV bands are provided. These are average values and are obtained using the consecutive annulus with a radius of 3 pixels starting from the center to the maximum radius till clumps are detected in the individual galaxy. $^{\star\star}$SDSS J1019 does not have UVIT NUV image, only total extinction values are given using GALEX FUV, NUV magnitudes.}
\resizebox{\textwidth}{!}{
\begin{tabular}{lcccc|ccc|cccccc}
\hline 
& \multicolumn{4}{c}{Total Av calculation with radius from col 2, Table \ref{magnitudes}} & \multicolumn{6}{c}{Max and min Av values for the radius till the SFCs are detected}\\
    \cline{2-5} \cline{6-11}
    &  & Internal & A$_{FUV}$ &A$_{NUV}$ & & A$_{FUV}$ & & &  A$_{NUV}$\\
     \cline{6-8} \cline{9-11}
%     &  & Internal & A$_{FUV}$ &A$_{NUV}$ &A$_{FUV_{max}}$   &A$_{FUV_{min}}$  & A$_{FUV_{mean}}$ & A$_{NUV_{max}}$   &A$_{NUV_{min}}$ &A$_{NUV_{mean}}$\\

Name   & $\beta$ & E(B-V)  & (mag) & (mag) & max & min &  mean  & max & min & mean\\
 (1) & (2) & (3) & (4) & (5) & (6) & (7) & (8) & (9) & (10) & (11)\\ \hline\hline
MRK 306  & -1.21 $\pm$ 0.01  & 0.299 $\pm$ 0.003 & 1.30 $\pm$ 0.01 & 1.03 $\pm$ 0.01 & 1.85 $\pm$ 0.04 &0.82 $\pm$ 0.21  & 1.29 $\pm$ 0.06& 1.47 $\pm$ 0.03  & 0.65 $\pm$ 0.17 & 1.03 $\pm$ 0.05  \\
MRK 721  & -0.92 $\pm$ 0.02 & 0.361 $\pm$ 0.003 & 1.50 $\pm$ 0.01 & 1.19 $\pm$ 0.01  & 2.27 $\pm$ 0.03 & 0.74 $\pm$ 0.19 & 1.55 $\pm$ 0.10 & 1.80 $\pm$ 0.02 & 0.59 $\pm$ 0.11 & 1.13 $\pm$ 0.09   \\
MRK 789   & -0.48 $\pm$ 0.01 & 0.456 $\pm$ 0.002 & 1.99 $\pm$ 0.01 & 1.58 $\pm$ 0.01 & 1.94 $\pm$ 0.01 & 1.64 $\pm$ 0.04 & 1.82 $\pm$ 0.04 & 1.54 $\pm$ 0.01 & 1.34 $\pm$ 0.03 & 1.45 $\pm$ 0.03   \\
NGC 3773  & -2.03 $\pm$ 0.13 & 0.123 $\pm$ 0.030 & 0.53 $\pm$ 0.13 & 0.44 $\pm$ 0.10 & 0.53 $\pm$ 0.23 & 0.28 $\pm$ 0.05 & 0.39 $\pm$ 0.05 & 0.42 $\pm$ 0.22 & 0.23 $\pm$  0.04 & 0.31 $\pm$ 0.04   \\
MRK 739  & -0.73 $\pm$ 0.01 & 0.402 $\pm$ 0.001 & 1.83 $\pm$ 0.01 & 1.40 $\pm$ 0.01 & 2.44 $\pm$ 0.01 & 1.59 $\pm$ 0.02 & 2.08$\pm$ 0.07 & 1.94 $\pm$ 0.01 & 1.27 $\pm$ 0.02 & 1.66 $\pm$ 0.06 \\
ESO509-IG066 & -0.84  $\pm$ 0.01 & 0.378  $\pm$ 0.002 & 1.65  $\pm$ 0.01& 1.31  $\pm$ 0.01 & 2.57 $\pm$ 0.02  & 1.23 $\pm$ 0.14 & 1.95 $\pm$ 0.10 & 2.04 $\pm$ 0.02 & 0.98 $\pm$ 0.19 & 1.55 $\pm$ 0.08    \\
SDSS J1436  & -0.20  $\pm$ 0.01 & 0.514  $\pm$ 0.004 & 2.23  $\pm$ 0.01 & 1.78  $\pm$ 0.01 & 2.32 $\pm$ 0.05 & 1.27  $\pm$ 0.16 & 1.83 $\pm$ 0.12 & 1.84 $\pm$ 0.04 & 1.01 $\pm$ 0.12 & 1.46  $\pm$ 0.10  \\
SDSS J1019$^{\star\star}$ & 0.76  $\pm$ 0.02 & 0.723  $\pm$ 0.004 & 3.25  $\pm$ 0.02 & 2.62  $\pm$ 0.01 & - & - & - &- & - &-\\
SDSS J1027 & 1.53  $\pm$ 0.01 & 0.886  $\pm$ 0.002 & 4.04  $\pm$ 0.01 & 3.10  $\pm$ 0.01 & 4.08 $\pm$ 0.03 & 3.13 $\pm$ 0.03 & 3.16 $\pm$ 0.16 & 3.13 $\pm$ 0.21 & 2.40 $\pm$ 0.02 & 2.76 $\pm$ 0.16  \\ \hline
\label{extinction}  
\end{tabular}}
\end{table*}

\begin{table*}
\caption{{Total SFR: Columns: (1) Name of the objects. (2) Observation wavebands. (3, 4) Extinction (external + internal) corrected flux density and luminosity. (5) Total star formation rate (SFR).}}
%\resizebox{\textwidth}{!}{
\begin{tabular}{lcccccccc}
\hline
Object & Band & Flux density  & Luminosity  & SFR  \\ 
 &  & (ergs s$^{-1}$ cm$^{-2}$ $\AA^{-1}$) & (ergs s$^{-1}$) & (M$_\odot$ yr$^{-1}$)\\
 (1) & (2) & (3) & (4) & (5) \\ \hline
MRK 306 & FUV& (5.28 $\pm$ 0.09) $\times 10^{-14}$ &(5.40 $\pm$ 0.09) $\times 10^{+43}$&4.36 $\pm$ 0.07\\
  & NUV&(2.46 $\pm$ 0.02) $\times 10^{-14}$& (3.83 $\pm$ 0.03) $\times 10^{+43}$ &3.09 $\pm$ 0.03\\ \hline
MRK 721 & FUV&(2.17 $\pm$ 0.05) $\times 10^{-14}$& (8.20 $\pm$ 0.20)$\times 10^{+43}$& 6.61 $\pm$ 0.15\\
  &NUV& (1.08 $\pm$ 0.01) $\times 10^{-14}$ & (6.24 $\pm$ 0.08) $\times 10^{+43}$& 5.04 $\pm$ 0.06\\ \hline
MRK 789& FUV&(3.63 $\pm$ 0.07) $\times 10^{-14}$& (1.31 $\pm$ 0.02) $\times 10^{+44}$ &10.58 $\pm$ 0.21\\
 & NUV& (1.94 $\pm$ 0.01) $\times 10^{-14}$ & (1.06 $\pm$ 0.01) $\times 10^{+44}$& 8.60 $\pm$ 0.08\\ \hline
NGC 3773& FUV& (6.74 $\pm$ 0.04) $\times 10^{-14}$& (4.34 $\pm$ 0.03) $\times 10^{+42}$ & 0.350 $\pm$ 0.002\\
 & NUV& (2.56 $\pm$ 0.01) $\times 10^{-14}$& (2.51 $\pm$ 0.01) $\times 10^{+42}$ & 0.203 $\pm$ 0.001\\ \hline
MRK 739& FUV& (3.77 $\pm$ 0.04)$\times 10^{-14}$& (1.22 $\pm$ 0.01) $\times 10^{+44}$& 9.91 $\pm$ 0.10\\
   & NUV& (4.84 $\pm$ 0.05) $\times 10^{-14}$& (1.45 $\pm$ 0.02) $\times 10^{+44}$& 11.71 $\pm$ 0.13\\ \hline
ESO509-IG066 (east) & FUV& (7.71 $\pm$ 0.15) $\times 10^{-15}$& (3.07 $\pm$ 0.06) $\times 10^{+43}$& 2.48 $\pm$ 0.04\\
& NUV& (4.17 $\pm$ 0.09) $\times 10^{-15}$& (2.53 $\pm$ 0.05) $\times 10^{+43}$& 2.04 $\pm$ 0.04\\ \hline
ESO509-IG066 (west) & FUV & (8.76 $\pm$ 0.16) $\times 10^{-15}$ & (3.49 $\pm$ 0.06) $\times 10^{+43}$ & 2.82 $\pm$ 0.05\\
& NUV& (4.46 $\pm$ 0.10) $\times 10^{-15}$& (2.70 $\pm$ 0.06) $\times 10^{+43}$& 2.18 $\pm$ 0.05\\ \hline
SDSS J1436&FUV& (6.89 $\pm$ 0.34) $\times 10^{-15}$& (6.07 $\pm$ 0.30) $\times 10^{+43}$& 4.90 $\pm$ 0.24\\
 & NUV& (4.11 $\pm$ 0.08) $\times 10^{-15}$& (5.51 $\pm$ 0.11) $\times 10^{+43}$& 4.45 $\pm$ 0.09\\ \hline
SDSS J1019 & FUV& (4.17 $\pm$ 0.07) $\times 10^{-14}$& (4.08 $\pm$ 0.07) $\times 10^{+44}$& 32.96 $\pm$ 0.62\\ \hline 
SDSS J1027 &FUV& (8.59 $\pm$ 0.05) $\times 10^{-15}$& (1.30 $\pm$ 0.07) $\times 10^{+44}$& 10.55 $\pm$ 0.62\\
  & NUV& (6.83 $\pm$ 0.07) $\times 10^{-15}$& (1.69 $\pm$ 0.02) $\times 10^{+44}$& 13.70 $\pm$ 0.15\\ \hline
\label{total_sfr}  
\end{tabular}
\end{table*}

\begin{table*}
\caption{{Stellar mass (M$_\star$) and specific star-formation rate (sSFR): Columns: (1) Name of the objects. (2, 3): WISE bands W1, W2 magnitudes centered at 3.368, 4.618 $\mu$m. (4) Stellar mass of the galaxies. (5) specific star-formation rate calculated using total SFR from column 5 of Table \ref{total_sfr} and stellar mass.}}
%\resizebox{\textwidth}{!}{
\begin{tabular}{lccccccccccc}
\hline
Object&W1 (mag) &W2 (mag) &M$_\star$ (M$_\odot$)&sSFR  (yr$^{-1}$) \\ 
(1) & (2)  & (3)  & (4)    & (5)\\ \hline
MRK 306&12.27 $\pm$ 0.02&12.01 $\pm$ 0.02& (2.24 $\pm$ 0.41) $\times 10^{9}$ & (2.07 $\pm$ 0.48) $\times 10^{-9}$\\
MRK 721&12.01 $\pm$ 0.02&11.82 $\pm$ 0.02& (1.64 $\pm$ 0.31) $\times 10^{10}$ & (4.48 $\pm$ 0.85) $\times 10^{-10}$\\
MRK 789&11.07 $\pm$ 0.02&10.67 $\pm$ 0.02& (1.04 $\pm$ 0.18) $\times 10^{10}$& (1.02 $\pm$ 0.18) $\times 10^{-9}$\\
NGC 3773&11.47 $\pm$ 0.02&11.36 $\pm$ 0.02& (6.98 $\pm$ 1.27) $\times 10^{8}$& (5.01 $\pm$ 0.92) $\times 10^{-10}$\\
MRK 739&9.91 $\pm$ 0.02&9.09 $\pm$ 0.02& (2.41 $\pm$ 0.42) $\times 10^{9}$& (4.11 $\pm$ 0.71) $\times 10^{-9}$\\
ESO509-IG066 east&12.14 $\pm$ 0.02&11.83 $\pm$ 0.02& (7.22 $\pm$ 1.34) $\times 10^{9}$& (3.43 $\pm$ 0.64) $\times 10^{-10}$\\
ESO509-IG066 west&11.05 $\pm$ 0.02&10.32 $\pm$ 0.02& (1.73 $\pm$ 0.30) $\times 10^{9}$& (2.03 $\pm$ 0.34) $\times 10^{-9}$\\
SDSS J1436&11.41 $\pm$ 0.02&11.42 $\pm$ 0.02& (2.05 $\pm$ 0.37) $\times 10^{11}$ & (2.39 $\pm$ 0.45) $\times 10^{-11}$\\
SDSS J1019&11.93 $\pm$ 0.02&11.84 $\pm$ 0.02& (8.67 $\pm$ 1.65) $\times 10^{10}$ & (3.46 $\pm$ 0.66) $\times 10^{-10}$\\
SDSS J1027&12.05 $\pm$ 0.04&11.89 $\pm$ 0.04& (7.94 $\pm$ 2.40) $\times 10^{10}$& (1.33 $\pm$ 0.40) $\times 10^{-10}$\\ \hline

\label{stellar_mass}
\end{tabular}
\end{table*}

\clearpage
\onecolumn
\begin{center}
\begin{longtable}{ccccccccc}
\caption{Surface density of SFR ($\Sigma_{SFR}$) of the SFCs: Columns: (1) The number of SFCs detected in the FUV and NUV images. (2, 3) RA and DEC of the central position of SFCs. (4, 5) semi-major and semi-minor axis in arcsec. (6) Area of the SFCs. (7) Position angle in radian. (8) The star formation rate density. $^{\star}$SFCs have AGN included which may have contributed to the UV emission (though not dominated according to Figure \ref{wise_color}). Hence, the true values can be equal or less than the quoted values. $^{\star\star}$SFCs 1 and 3 of SDSS J1019 are masked (radius: 3 pixels) to minimize the AGN where SFC 3 gets fully masked and provides NAN value.} \label{tab:long} \\

\hline SFC No&RA (J2000)&Dec(J2000)&semi-major &semi-minor&Area&P.A&$\Sigma_{SFR}$\\
  & (hh:mm:ss) & (dd:mm:ss) & ($\arcsec$) & ($\arcsec$) &(kpc$^2)$ & (radian) & (M$_\odot$ yr$^{-1}$ kpc$^{-2}$) \\
  (1) & (2) & (3) & (4) & (5) & (6)  & (7) & (8)
% \hline \multicolumn{1}{|c|}{\textbf{} & \multicolumn{1}{c|}{\textbf{Second column}} & \multicolumn{1}{c|}{\textbf{Third column}} & \multicolumn{1}{c|}{\textbf{Third column}}
\\ \hline 
\endfirsthead

\multicolumn{3}{c}%
{{\bfseries \tablename\ \thetable{} -- continued from previous page}} \\
\hline SFC No&RA (J2000)&Dec(J2000)&semi-major &semi-minor&Area&P.A&$\Sigma_{SFR}$\\
  & (hh:mm:ss) & (dd:mm:ss) & ($\arcsec$) & ($\arcsec$) &(kpc$^2)$ & (radian) & (M$_\odot$ yr$^{-1}$ kpc$^{-2}$)  \\
(1) & (2) & (3) & (4) & (5) & (6)  & (7) & (8)
\\ \hline \endhead

\hline \multicolumn{4}{c}{{Continued on next page......}} \\ \hline \endfoot

\hline \hline \endlastfoot
{\textcolor{blue}{MRK 306 FUV}}\\
%1&22:31:49.30&19:41:46.17&1.31&0.63&0.3&-0.22&0.0747$\pm$0.0174\\
1&22:31:51.83&19:41:36.83&0.97&0.58&0.20&-0.67&0.0842 $\pm$ 0.0229\\
2&22:31:51.29&19:41:28.40&1.46&0.98&0.52&-0.94&0.1442 $\pm$ 0.0185\\
3&22:31:51.15&19:41:27.65&1.46&0.60&0.32&0.11&0.1471 $\pm$ 0.0237\\
4&22:31:51.33&19:41:24.76&0.75&0.57&0.16&-1.40&0.1544 $\pm$ 0.0349\\
5&22:31:51.19&19:41:25.15&2.48&1.07&0.96&-0.92&0.1288 $\pm$ 0.0128\\
6&22:31:51.14&19:41:37.88&1.33&1.04&0.50&-1.04&0.1176 $\pm$ 0.0170\\
7&22:31:51.34&19:41:42.34&3.46&1.89&2.37&0.77&0.1156 $\pm$ 0.0077\\
8&22:31:51.18&19:41:33.48&2.90&2.32&2.44&-1.47&0.1596 $\pm$ 0.0090\\
9&22:31:51.16&19:41:20.88&3.29&1.42&1.70&-1.15&0.0584 $\pm$ 0.0065\\
{\textcolor{blue}{MRK 306 NUV}}\\
1&22:31:52.39&19:41:49.45&0.78&0.63&0.18&-1.00&0.0423 $\pm$ 0.0100\\
2&22:31:52.25&19:41:49.70&1.62&0.68&0.4&1.50&0.0384 $\pm$ 0.0064\\
3&22:31:52.45&19:41:52.24&1.92&0.88&0.61&-0.54&0.0249 $\pm$ 0.0041\\
%4&22:31:49.31&19:41:45.90&2.0&1.27&0.92&-1.27&0.0826$\pm$0.0061\\
4&22:31:52.46&19:41:38.80&2.24&1.27&1.03&-1.11&0.0431 $\pm$ 0.0042\\
5&22:31:49.88&19:41:25.01&1.78&1.12&0.72&-0.32&0.0609 $\pm$ 0.0060\\
6&22:31:50.28&19:41:21.34&0.96&0.54&0.19&0.20&0.0437 $\pm$ 0.0099\\
7&22:31:51.23&19:41:33.91&2.70&2.40&2.35&0.43&0.1957 $\pm$ 0.0059\\
8&22:31:51.28&19:41:27.64&4.30&3.41&5.33&1.00&0.1169 $\pm$ 0.0030\\
9&22:31:51.30&19:41:42.32&5.76&2.92&6.11&0.71&0.0963 $\pm$ 0.0026\\
10&22:31:51.11&19:41:39.20&1.30&0.71&0.34&1.33&0.1276 $\pm$ 0.0127\\
11&22:31:51.21&19:41:21.75&2.24&1.36&1.11&-1.11&0.0684 $\pm$ 0.0051\\
12&22:31:50.95&19:41:32.83&2.06&0.96&0.72&0.60&0.0652 $\pm$ 0.0062\\
13&22:31:51.64&19:41:35.87&2.08&1.57&1.19&-1.52&0.0447 $\pm$ 0.0040\\
14&22:31:51.83&19:41:36.40&4.26&1.66&2.57&-1.57&0.0456 $\pm$ 0.0027\\
15&22:31:50.89&19:41:25.88&1.68&1.38&0.84&0.58&0.0400 $\pm$ 0.0045\\
16&22:31:50.93&19:41:18.85&2.14&1.90&1.48&-1.15&0.0446 $\pm$ 0.0036\\
17&22:31:50.78&19:41:37.85&2.55&1.68&1.56&0.60&0.0337 $\pm$ 0.0030\\
18&22:31:50.51&19:41:16.13&4.06&1.35&1.99&0.16&0.0250 $\pm$ 0.0023\\
19&22:31:51.90&19:41:43.37&2.45&0.73&0.65&-0.44&0.0269 $\pm$ 0.0042\\ \hline
{\textcolor{blue}{MRK 721 FUV}}\\
1&10:23:32.66&10:57:35.44&3.11&1.67&6.48&0.85&0.1072 $\pm$ 0.0078\\
2&10:23:32.64&10:57:31.89&1.52&1.13&2.14&0.36&0.1154 $\pm$ 0.0141\\
3&10:23:32.53&10:57:40.31&4.37&2.98&16.24&-0.52&0.0915 $\pm$ 0.0046\\
4&10:23:32.86&10:57:34.13&1.66&0.66&1.37&1.12&0.0759 $\pm$ 0.0144\\
5&10:23:32.92&10:57:30.35&2.15&1.24&3.32&0.0&0.0452 $\pm$ 0.0071\\
{\textcolor{blue}{MRK 721 NUV}}\\
1&10:23:32.53&10:57:39.06&5.70&3.07&21.82&-0.82&0.0985 $\pm$ 0.0022\\
2&10:23:32.63&10:57:35.01&1.16&0.67&0.97&0.16&0.1898 $\pm$ 0.0143\\
3&10:23:32.58&10:57:31.57&3.39&2.0&8.45&-0.09&0.1058 $\pm$ 0.0036\\
4&10:23:32.73&10:57:29.39&3.35&1.30&5.43&0.38&0.0070 $\pm$ 0.0037\\
5&10:23:32.99&10:57:36.64&3.21&1.1&4.40&0.41&0.0662 $\pm$ 0.0040\\
6&10:23:33.12&10:57:43.12&3.57&3.17&14.11&0.34&0.0358 $\pm$ 0.0016\\
7&10:23:32.80&10:57:48.40&2.50&1.81&5.64&-0.12&0.0196 $\pm$ 0.0019\\ \hline
{\textcolor{blue}{MRK 789 FUV}}\\
1&13:32:24.22&11:06:23.37&1.66&1.25&2.52&0.50&0.3974 $\pm$ 0.0268\\
2$^{\star}$&13:32:24.01&11:06:20.19&2.61&2.16&6.83&0.88&$\leq$0.6596 $\pm$ 0.0210\\
{\textcolor{blue}{MRK 789 NUV}}\\
1&13:32:24.22&11:06:23.31&1.66&1.60&3.21&0.27&0.4973 $\pm$ 0.0136\\
2&13:32:24.00$^{\star}$&11:06:20.00&3.20&2.92&11.28&0.56&$\leq$0.6025 $\pm$ 0.0080\\ \hline
{\textcolor{blue}{NGC 3773 FUV}}\\
1&11:38:13.00&12:06:44.37&3.78&2.86&0.26&0.38&0.9069 $\pm$ 0.0079\\
{\textcolor{blue}{NGC 3773 NUV}}\\
1&11:38:12.99&12:06:44.19&4.73&3.82&0.44&0.4&0.4869 $\pm$ 0.0026\\
{\textcolor{blue}{MRK 739 FUV}}\\
1&11:36:28.22&21:35:31.15&2.64&2.23&6.44&0.76&0.0589 $\pm$ 0.0026\\
2&11:36:27.82&21:35:30.65&2.48&2.30&6.24&-1.56&0.1055 $\pm$ 0.0035\\
3&11:36:28.16&21:35:35.94&1.36&1.05&1.56&-0.54&0.0569 $\pm$ 0.0052\\
4&11:36:27.42&21:35:36.86&2.67&1.22&3.55&0.30&0.0504 $\pm$ 0.0032\\
5&11:36:27.22&21:35:32.79&2.17&0.71&1.68&1.34&0.0517 $\pm$ 0.0047\\
6&11:36:27.30&21:35:29.25&4.10&2.41&10.8&1.43&0.0444 $\pm$ 0.0017\\
7&11:36:28.08&21:35:40.03&3.21&2.60&9.11&-1.10&0.0457 $\pm$ 0.0019\\
8&11:36:27.76&21:35:37.20&2.66&1.57&4.58&-0.41&0.0356 $\pm$ 0.0024\\
9&11:36:27.61&21:35:24.32&2.71&1.47&4.35&-0.42&0.0319 $\pm$ 0.0023\\
10&11:36:27.89&21:35:24.37&2.25&1.36&3.35&1.03&0.0279 $\pm$ 0.0025\\
11&11:36:28.64&21:35:31.90&1.95&1.01&2.16&1.27&0.0281 $\pm$ 0.0031\\
12&11:36:28.37&21:35:26.22&1.90&0.98&2.03&0.89&0.0217 $\pm$ 0.0028\\
13&11:36:28.19&21:35:25.46&1.75&1.25&2.39&1.41&0.0247 $\pm$ 0.0028\\
14&11:36:27.72&21:35:41.84&2.70&1.00&2.97&-0.23&0.0204 $\pm$ 0.0022\\
{\textcolor{blue}{MRK 739 NUV}}\\
1&11:36:29.34&21:35:46.22&1.93&1.74&3.67&-0.55&0.1877 $\pm$ 0.0105\\
2&11:36:28.93&21:35:46.59&2.98&2.03&6.62&-1.44&0.2538 $\pm$ 0.0091\\
3&11:36:29.01&21:35:52.55&2.22&0.82&1.99&-0.86&0.1191 $\pm$ 0.0114\\
4&11:36:28.58&21:35:52.94&4.20&1.99&9.14&0.29&0.1227 $\pm$ 0.0054\\
5&11:36:29.33&21:35:53.78&4.15&2.46&11.16&-1.27&0.1447 $\pm$ 0.0053\\
6&11:36:28.30&21:35:50.40&1.23&0.56&0.75&-0.99&0.1608 $\pm$ 0.0215\\
7&11:36:28.44&21:35:44.41&5.36&2.94&17.23&-0.92&0.1021 $\pm$ 0.0036\\
8&11:36:29.65&21:35:43.64&3.32&1.01&3.67&1.00&0.0768 $\pm$ 0.0067\\
9&11:36:29.32&21:35:39.09&3.46&1.45&5.49&-1.02&0.0468 $\pm$ 0.0043\\ \hline
{\textcolor{blue}{ESO509-IG066 FUV}}\\
1&13:34:42.80&-23:26:49.28&1.81&0.59&1.42&-0.75&0.0072 $\pm$ 0.0006\\
2&13:34:41.90&-23:26:50.60&4.04&2.85&15.28&0.57&0.0116 $\pm$ 0.0002\\
3&13:34:43.05&-23:26:47.13&1.74&1.14&2.64&0.94&0.0089 $\pm$ 0.0005\\
4&13:34:43.11&-23:26:44.73&1.16&0.68&1.05&1.10&0.0082 $\pm$ 0.0007\\
5&13:34:43.21&-23:26:43.56&2.08&0.98&2.71&-1.12&0.0064 $\pm$ 0.0004\\
6&13:34:42.84&-23:26:40.81&1.87&1.14&2.82&0.60&0.0071 $\pm$ 0.0004\\
7&13:34:43.13&-23:26:35.07&1.25&0.62&1.02&-0.40&0.0066 $\pm$ 0.0007\\ 
{\textcolor{blue}{ESO509-IG066 NUV}}\\
1&13:34:42.30&-23:26:53.88&2.77&2.20&8.08&-1.31&0.0414 $\pm$ 0.0022\\
2&13:34:42.53&-23:26:52.64&1.78&1.20&2.83&-1.43&0.0549 $\pm$ 0.0042\\
3&13:34:42.12&-23:26:58.91&1.45&1.00&1.93&-0.02&0.0303 $\pm$ 0.0038\\
4&13:34:41.95&-23:26:55.05&1.79&1.59&3.76&0.99&0.0318 $\pm$ 0.0028\\
5&13:34:43.58&-23:26:45.78&1.92&0.83&2.11&-1.09&0.0400 $\pm$ 0.0042\\
6&13:34:43.41&-23:26:43.82&2.01&0.74&1.99&0.08&0.0358 $\pm$ 0.0041\\
7&13:34:43.45&-23:26:51.23&2.27&1.63&4.91&-0.18&0.0312 $\pm$ 0.0024\\
8&13:34:43.26&-23:26:45.91&1.12&0.71&1.06&0.41&0.0308 $\pm$ 0.0052\\
9&13:34:43.81&-23:26:40.21&0.73&0.56&0.54&0.95&0.0334 $\pm$ 0.0075\\ \hline
{\textcolor{blue}{SDSS J1436 NUV}}\\
1&14:36:48.36&18:20:47.55&0.76&0.61&1.29&-1.5&0.0428 $\pm$ 0.0100\\
2&14:36:47.85&18:20:42.53&1.42&0.82&3.23&-1.44&0.0681 $\pm$ 0.0080\\
3&14:36:48.01&18:20:45.00&1.30&0.98&3.54&1.45&0.0504 $\pm$ 0.0066\\
4&14:36:47.89&18:20:32.94&1.28&0.59&2.10&0.75&0.0387 $\pm$ 0.0075\\
5&14:36:48.16&18:20:37.66&1.09&1.00&3.03&0.19&0.0484 $\pm$ 0.0070\\ \hline
{\textcolor{blue}{SDSS J1019 FUV$^{\star\star}$}}\\
1&10:19:20.82&49:07:01.42&1.77&0.99&5.73&-1.31&0.0640 $\pm$ 0.0115\\
2&10:19:20.53&49:07:00.70&2.56&1.45&12.13&-0.98&0.5434 $\pm$ 0.0230\\
3&10:19:21.03&49:07:02.96&0.84&0.68&1.87&0.48&NAN&\\
4&10:19:21.34&49:07:04.00&3.43&2.31&25.9&0.58&0.1806 $\pm$ 0.0091\\
5&10:19:21.40&49:07:08.16&1.67&1.46&7.97&-0.37&0.1790 $\pm$ 0.0163\\
%6&10:19:22.12&49:07:10.58&3.56&2.1&24.44&0.05&NAN&\\ \hline

{\textcolor{blue}{SDSS J1027 FUV}}\\
1$^{\star}$&10:27:00.53&17:49:00.97&0.81&0.40&1.59&-1.25&$\leq$0.2663 $\pm$ 0.079\\
{\textcolor{blue}{SDSS J1027 NUV}}\\
1&10:27:00.75&17:48:59.91&1.17&0.85&4.82&-0.44&0.2148 $\pm$ 0.0111\\
2&10:27:00.49&17:49:01.00&3.76&2.34&42.89&-0.28&0.1852 $\pm$ 0.0034\\
3$^{\star}$&10:27:00.79&17:49:01.92&1.53&1.20&8.96&0.22&$\leq$0.1193 $\pm$ 0.006\\

\label{sfc_sfr}
\end{longtable}
\end{center}
\clearpage
\twocolumn

\appendix 
\section{Appendix}
Here we present some additional images to support some of the analysis and arguments in the main text with proper references.  

\begin{figure*}
\includegraphics[width=\columnwidth, trim= 0 0 0 0]{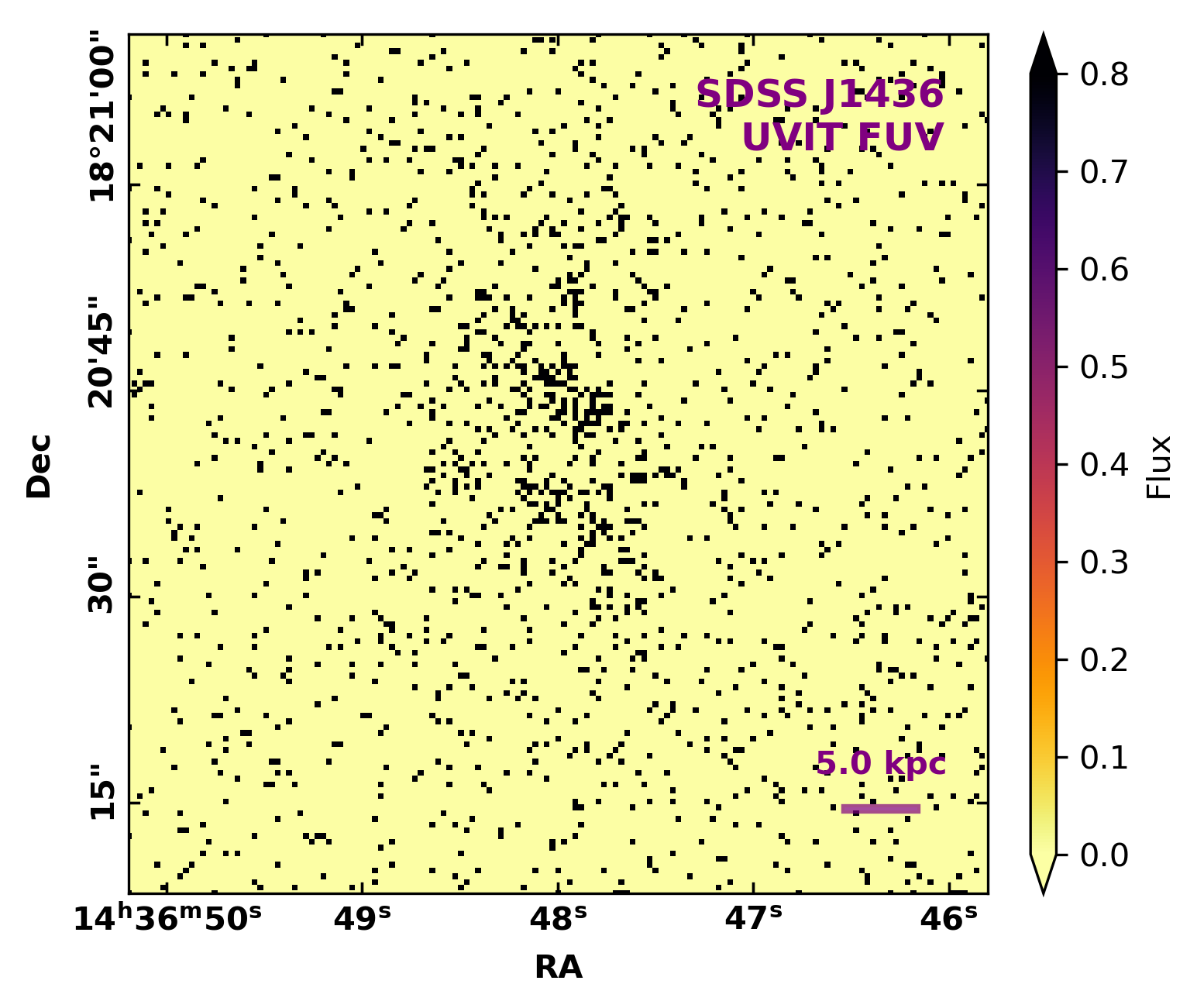}\includegraphics[width=\columnwidth, trim= 0 0 0 0]{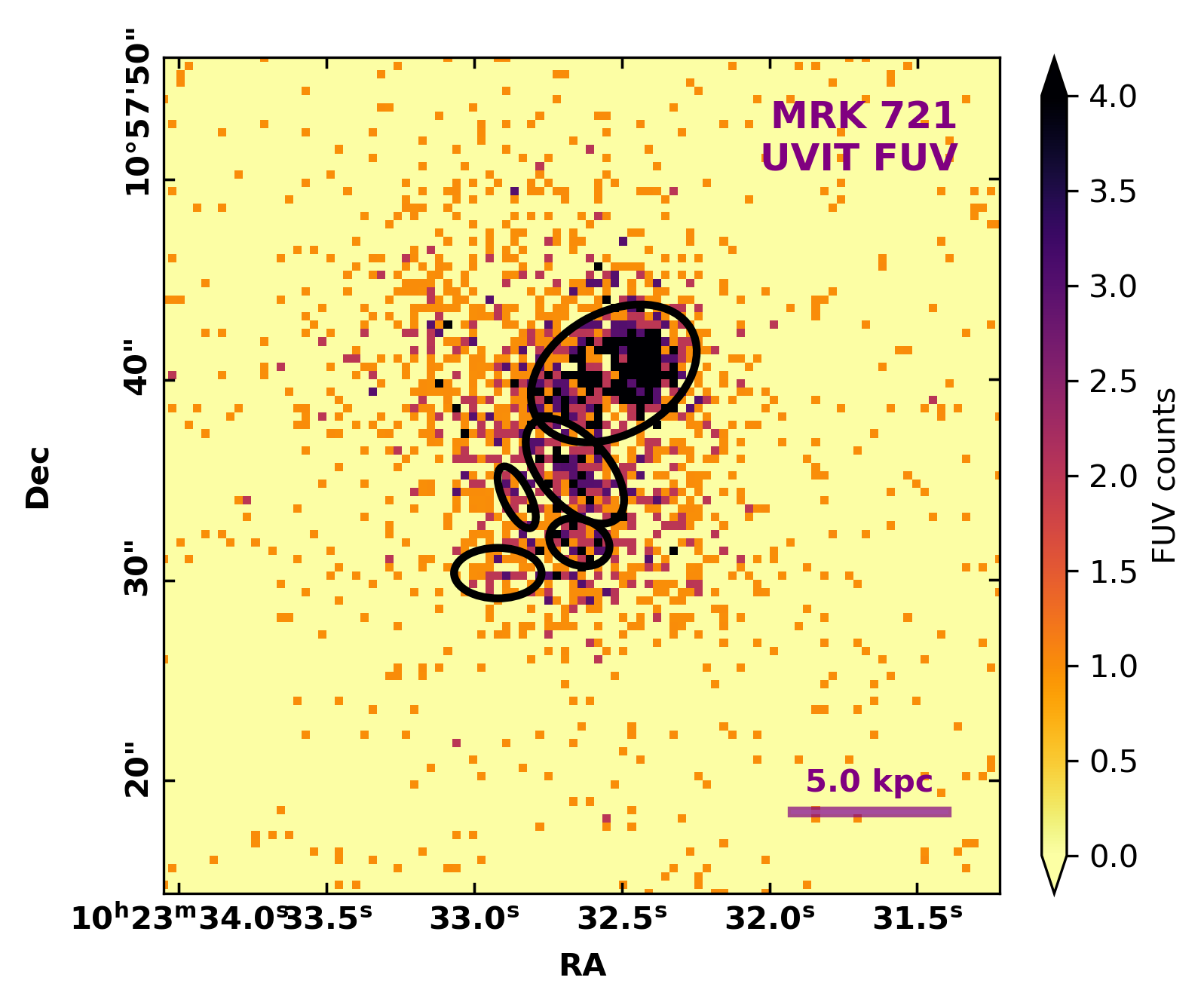}
\caption{\small The unsmoothed UVIT images. ({\it left}) The FUV image of SDSS J1436 where {\it sextractor} could not detect any SFCs. ({\it right}) The FUV image of MRK 721. The smoothed images may look like some of the SFCs do not like convincing while other non-detected regions look like SFCs. However, it depends on the three-step criteria as mentioned in section \ref{id_SFCs}.}
\label{unsmoothed_fig}
\end{figure*}

% \bibliographystyle{mnras}
% \bibliography{ms}

\label{lastpage}
\end{document}